\documentclass[preprint,12pt]{aastex} 
\usepackage{natbib}

\newcommand{\asec}{\ifmmode^{\prime\prime}\else$^{\prime\prime}$\fi}

\newcommand{\bq}{\begin{equation}}
\newcommand{\eq}{\end{equation}}

\newcommand{\p}{~$\pm$~}

\newcommand{\3}{$_3$}
\newcommand{\1}{$^{-1}$}
\newcommand{\2}{$_2$}

\newcommand{\simgt}{\lower.5ex\hbox{$\; \buildrel > \over \sim \;$}}
\newcommand{\simlt}{\lower.5ex\hbox{$\; \buildrel < \over \sim \;$}}
\newcommand{\kms}{km~s$^{-1}$}
\newcommand{\dvi}{\Delta \rm{v}_{int}}

\newcommand{\dvo}{\Delta \rm{v}_{obs}}
\newcommand{\dvoa}{\Delta \rm{v}_{obs}(1,1)}

\newcommand{\tm}{\tau_{\rm{m}}}
\newcommand{\ta}{\tm(1,1)}

\newcommand{\tc}{\tm(3,3)}

\begin{document}

\title{The Nature of the Molecular Environment\\ within 5~pc of the
Galactic Center}
 
\author{
Robeson M. Herrnstein\altaffilmark{1} and
Paul T.P. Ho\altaffilmark{2}}

\altaffiltext{1}{Department of Astronomy, Columbia University, 550 West 120th Street, Mail Code 5246, New York, NY 10027, herrnstein@astro.columbia.edu}
\altaffiltext{1}{Harvard-Smithsonian Center for Astrophysics, 60 Garden Street, Cambridge, MA 02138, \\pho@cfa.harvard.edu}

\slugcomment{Accepted for publication in ApJ}

\begin{abstract}
We present a detailed study of molecular gas in the central 10~pc of
the Galaxy through spectral line observations of four rotation
inversion transitions of NH\3 made with the Very Large Array.  Updated
line widths and NH\3(1,1) opacities are presented, and temperatures,
column densities, and masses are derived for the major molecular
features.  We examine the impact of Sgr A East on molecular material
at the Galactic center and find that there is no evidence that the
expansion of this shell has moved a significant amount of the 50
km~s$^{-1}$ giant molecular cloud.  The western streamer, however,
shows strong indications that it is composed of material swept-up by
the expansion of Sgr A East.  Using the mass and kinematics of the
western streamer, we calculate an energy of
$E_{SN}=(2-9)\times10^{51}$~ergs for the progenitor explosion and
conclude that Sgr A East was most likely produced by a single
supernova.  The temperature structure of molecular gas in the central
$\sim20$~pc is also analyzed in detail.  We find that molecular gas
has a ``two-temperature'' structure similar to that measured by
\citet{hut93b} on larger scales.  The largest observed line ratios,
however, cannot be understood in terms of a two-temperature model, and
most likely result from absorption of NH\3(3,3) emission by cool
surface layers of clouds.  By comparing the observed NH\3
(6,6)-to-(3,3) line ratios, we disentangle three distinct molecular
features within a projected distance of 2~pc from Sgr~A*.  Gas
associated with the highest line ratios shows kinematic signatures of
both rotation and expansion.  The southern streamer shows no
significant velocity gradients and does not appear to be directly
associated with either the circumnuclear disk or the nucleus.  The
paper concludes with a discussion of the line-of-sight arrangement of
the main features in the central 10~pc.
\end{abstract}

\keywords{Galaxy: center --- ISM: clouds --- ISM: molecules --- radio
lines: ISM}

\section{Introduction \label{intro}}

Since its discovery by \citet{balick74}, the bright ($\sim1$~Jy),
compact radio source called Sagittarius A* (Sgr~A*) has been
hypothesized to mark the nucleus of the Milky Way.  Today, the
dynamical center of the Galaxy has been precisely determined through
infrared observations of stellar orbits in the central $1''$ (0.04~pc)
\citep{sch02,ghe03}.  The inferred mass of $\sim4\times10^6$~$M_\odot$
contained within a radius of $\sim60$~AU is some of the most
convincing evidence for the existence of a supermassive black hole.
Based on Very Long Baseline Array observations of the proper motion of
Sgr~A*, \citet{rei04} find that the peculiar motion of Sgr~A* relative
to the dynamical center of the Galaxy is $-0.4\pm0.9$ km~s$^{-1}$.
Such a small velocity so near a supermassive black hole implies that
Sgr~A* must contain at least $10^6$~$M_\odot$ \citep{rei04}.  It is
therefore most likely that that the radio emission from Sgr~A*
originates in either the inner region of a radiatively inefficient
accretion flow \citep{quataert03} or the nozzle of a jet emanating
from the supermassive black hole \citep{yuan02}.

With a bolometric luminosity of $\sim10^{36}$~ergs~s$^{-1}$
\citep[][and references therein]{narayan98}, the emission from Sgr~A*
is extremely sub-Eddington ($L_{Edd}=5\times10^{44}$~ergs~s$^{-1}$ for
a $4\times10^6~M_\odot$ black hole).  However, the close proximity of
Sgr~A* ($d=8.0\pm0.5$~kpc, \citet{rei93}), makes it uniquely suited
for the detailed study of the environment around a supermassive black
hole.  In the central $\sim2$~pc of the Galaxy, Sgr~A* is surrounded
by multiple arcs of ionized gas, which together are called the
``mini-spiral'' or Sgr A West \citep{ekers83,lo83,rob93}.  These arcs
are surrounded by an apparent ``ring'' of dense ($\sim10^5~$cm$^{-3}$)
molecular material called the circumnuclear disk (CND,
\citet{gus87,wri01}).  Neither of these features appears to be
gravitationally stable, and it is estimated that (barring other
external pressures) this material will accrete into the nucleus on the
order of $10^4-10^5$~yr \citep{wri01}.

For the past two decades, the origin of the dense gas and dust in the
CND and Sgr A West has remained unclear. Many attempts have been made
to detect kinematic connections between two nearby giant molecular
clouds (GMCs) and the CND. This work has focused primarily on
observations of rotation inversion transitions of NH\3 at frequencies
of 23--25~GHz.  These transitions trace densities similar to those
traced by the 3~mm HCN(1-0) transition ($n_{H_2}\sim10^5$cm$^{-3}$),
but the higher energies and lower opacities of the NH\3 transitions
significantly reduce the effects of self-absorption and absorption
along the line-of-sight (LOS) \citep{mcg01}.  Due to the close
frequency spacing of these spectral lines, observations of multiple
transitions can be made with the same instrument and receiver making
calculation of line ratios (and therefore rotational temperatures)
relatively straightforward.  In addition, hyperfine splitting of the
NH\3 lines enables direct calculation of gas opacity \citep{ho83}.

Previous NH\3 observations detect a long filamentary ``southern
streamer'' extending northwards from the ``20 km s\1'' GMC
(M--0.13--0.08; \citet{gus81}) towards the the southeastern edge of
the CND \citep{oku89,ho91,coi99,coi00}.  (See Figure \ref{fig:f1}c for
positions of the main molecular features in the region.)  Increased
line widths and evidence of heating as the southern streamer
approaches the Galactic center indicate that gas may be flowing from
the 20 km s\1 GMC towards the circumnuclear region.  Supporting
morphological evidence for this connection has come from observations
of HCN(3--2) \citep{mar95}, $^{13}$CO(2--1) \citep{zyl90}, and 1~mm
continuum emission \citep{den93,zyl98}.  However, the southern
streamer shows neither the velocity gradient nor curved morphology
that would be expected for a cloud infalling towards the nucleus
\citep{coi99}.  NH\3 emission also becomes very faint near the CND
making it difficult to study the gas in this region \citep{mcg01}.
Two additional connections originating in the northern fork of the 20
km~s$^{-1}$ GMC \citep{coi99,coi00} and the northern ridge
\citep{mcg01} have also been suggested based on NH\3 observations.  A
fourth possible connection originating in the ``50 km~s$^{-1}$'' GMC
(M--0.03--0.07) was detected in HCN(1--0), but it has not been
observed in NH\3 \citep{ho95}.

In this paper, we combine data from previous interferometric
observations of four rotation inversion transitions of NH\3 to study
the molecular environment in this region in detail.  The observations
and velocity integrated maps, which were previously published in
\citet{mcg01} and \citet{her02}, are briefly reviewed in \S \ref{obs}.
We then derive physical parameters for the molecular clouds in \S
\ref{derparm}, and use these results to investigate the impact of Sgr
A East on molecular gas in the central 10~pc in \S \ref{sgeast}.
Based on the interaction of Sgr A East with the western streamer, we
place limits on the energy of the progenitor explosion.  In \S
\ref{trots}, we discuss the detailed temperature structure of
molecular clouds at the Galactic center.  Measured temperatures for
clouds more than 2~pc from Sgr A* are consistent with a
two-temperature structure in which roughly one quarter of the gas is
contained in a hot ($\sim200$~K) component.  This structure is similar
to the model used by \citet{hut93b} to characterize temperatures on
larger scales.  NH\3 (6,6)-to-(3,3) line ratios in the central 2~pc,
however, exceed theoretical limits for gas in LTE \citep{her02}.  In
\S \ref{un-phys}, we model these line ratios as resulting from
absorption of NH\3(3,3) by cool material in a shielded layer of the
cloud.  In \S \ref{hlrllr}, we disentangle emission from three
distinct molecular clouds within a projected distance of 2~pc from
Sgr~A* through a comparison of NH\3(3,3) and (6,6) emission.  We
conclude this paper with a model of the line-of-sight (LOS)
distribution of the main features in the central 10~pc based on our
molecular data and in the context of existing data at other
wavelengths.

\section{Observations and Velocity Integrated Maps \label{obs}}

The large range of velocities observed at the Galactic center and the
relatively narrow velocity coverage of current interferometers at
centimeter wavelengths make spectral line observations of the region
difficult.  A velocity coverage of $\pm110$~\kms ~is necessary to
detect all of the emission from the CND, but additional clouds with
velocities as high as --185~\kms ~have been detected near Sgr~A*
\citep{zha95}.  Using the Very Large Array (VLA)\footnote{The National
Radio Astronomy Observatory is a facility of the National Science
Foundation operated under cooperative agreement by Associated
Universities, Inc.}, we observed the central 10~pc of the Galaxy in
NH\3(1,1), (2,2), and (3,3) in 1999 March.  The five pointing mosaic
fully samples the central $5'$ (10~pc) of the Galaxy with a velocity
coverage of of --140 to +130~\kms.  The resulting data are more
complete in both velocity coverage and spatial sampling than previous
interferometric observations of NH\3 emission in this region, which
tended to concentrate on velocities associated with the GMCs
\citep{coi99}.  However, the wide velocity coverage of our data is
gained at the expense of channel width, which is only 9.8~km~s$^{-1}$.

\begin{figure}
\includegraphics[width=6.0in,angle=270]{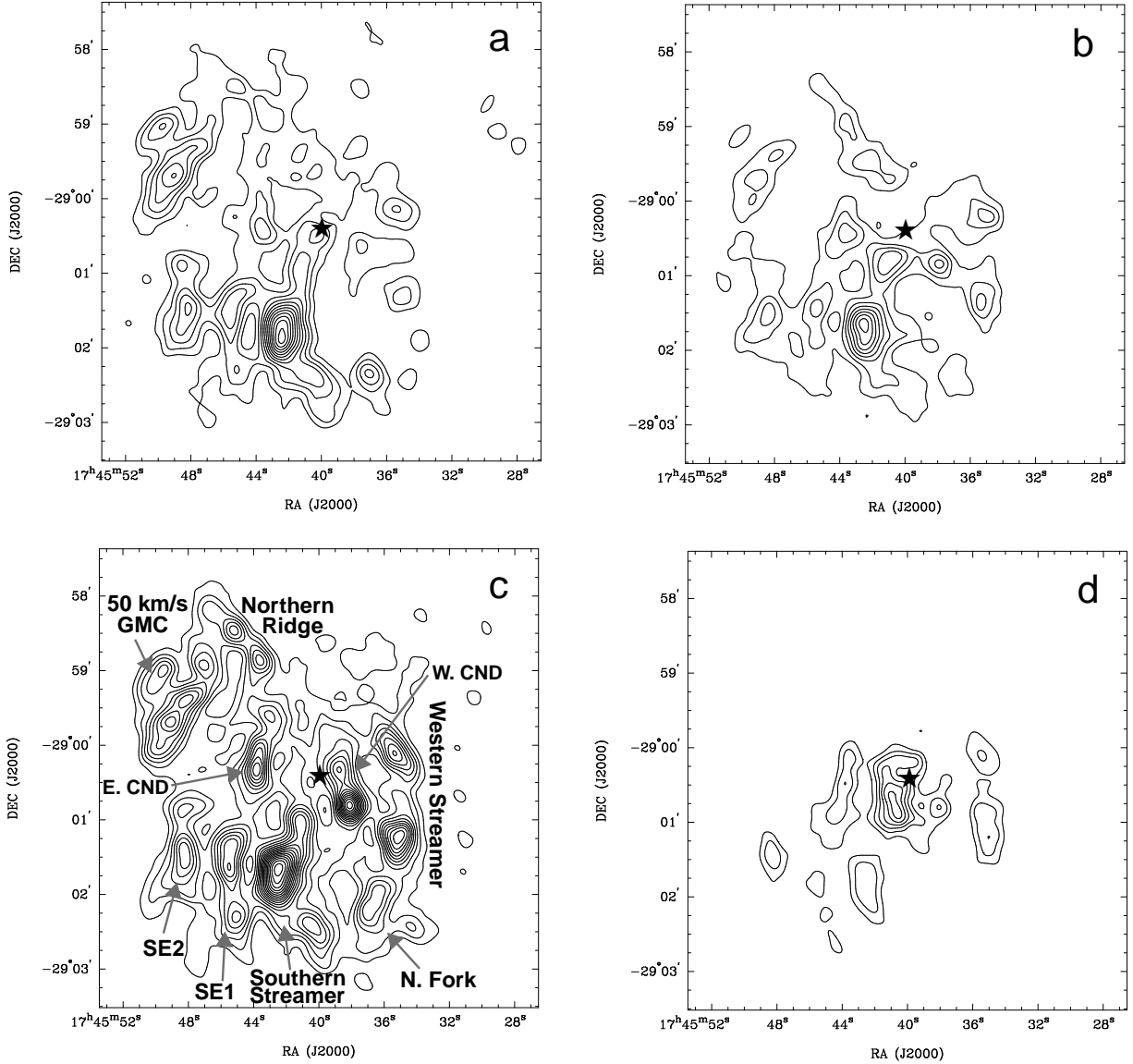}  
\caption{ \label{fig:f1} Velocity integrated maps of {\it a)}
NH\3(1,1), {\it b)} NH\3(2,2), {\it c)} NH\3(3,3), and {\it d)}
NH\3(6,6).  Contours are in steps of 1.32~Jy~beam$^{-1}$~km~s$^{-1}$
in all four panels.  The rms noise level, $\sigma_{JK}$, for each
$(J,K)$ rotation inversion transition is
$\sigma_{11}=0.28$~Jy~beam$^{-1}$~km~s$^{-1}$,
$\sigma_{22}=0.30$~Jy~beam$^{-1}$~km~s$^{-1}$,
$\sigma_{33}=0.33$~Jy~beam$^{-1}$~km~s$^{-1}$, and
$\sigma_{66}=0.33$~Jy~beam$^{-1}$~km~s$^{-1}$ \citep{mcg01,her02}.
The position of Sgr A* is labeled by a star in each panel.  The main
molecular features in the central 10~pc are labeled in panel {\it c}.}
\end{figure}

Figure \ref{fig:f1}a-c shows velocity integrated maps of NH\3(1,1),
(2,2), and (3,3) emission from the central 10~pc (see \citet{mcg01}
for details on the observational setup and data reduction).  The
position of Sgr~A*
($\alpha_{2000}=17^h45^m40^s.0,~\delta_{2000}=-29^\circ00'26''.6$) is
marked by a star in each panel.  Major molecular features discussed in
this paper are labeled in Figure \ref{fig:f1}c.  After application of
a Gaussian taper to de-emphasize long baselines, the synthesized beam
in all three maps is roughly $16''\times14''$ with a position angle of
$\sim0^\circ$.  The rms noise level, $\sigma_{JK}$, for each $(J,K)$
rotation inversion transition is
$\sigma_{11}=0.28$~Jy~beam$^{-1}$~km~s$^{-1}$,
$\sigma_{22}=0.30$~Jy~beam$^{-1}$~km~s$^{-1}$, and
$\sigma_{33}=0.33$~Jy~beam$^{-1}$~km~s$^{-1}$ \citep{mcg01}.  Contours
in all four panels of Figure \ref{fig:f1} are in steps of
1.32~Jy~Beam\1~\kms, thereby facilitating comparisons between
different transitions.  However, all emission exceeding $3\sigma_{JK}$
is considered significant (see \citet{mcg01} for maps showing
$3\sigma_{JK}$ contours). In order to have a constant noise level
across the map, we have not corrected for primary beam attenuation at
the edge of our mosaic, and the data are not sensitive to emission
$\simgt150''$ from Sgr~A* \citep{mcg01}.  The 50 km~s$^{-1}$ GMC,
molecular ridge (composed of SE1 and SE2), and southern streamer all
extend past the edge of the mosaic \citep[see][for large-scale maps
showing the extent of these clouds]{arm85,den93,zyl98}. However, the
outer edges of both the northern ridge and western streamer occur well
within our mosaic and reflect intrinsic cloud edges \citep{mcg01}.

As reported in \citet{mcg01}, we find dense molecular gas throughout
much of the region.  A notable exception is the far western and
northwestern parts of the mosaic.  These regions also lack emission
from HCN(1--0) \citep{wri01} and thermal dust
\citep{den93,zyl98,pierce00}.  Emission from NH\3 is predominately
associated with the 20 and 50 km~s$^{-1}$ GMCs.  Both the southern
streamer and northern fork \citep{coi99} are extensions of the 20
km~s$^{-1}$ GMC.  The two southeastern clouds (SE1 and SE2) form part
of the ``molecular ridge'' that connects this GMC to the 50
km~s$^{-1}$ GMC in the northeast \citep{coi00}.  Only the CND,
northern ridge, and western streamer show velocities outside the
$\sim20-50$~km~s$^{-1}$ range, and are not direct extensions of these
GMCs.  Overall, we find some evidence for kinematic connections
between giant molecular clouds in the central 10~pc, but the expansion
of the Sgr~A~East shell appears to be a dominating force in the
region, possibly pushing much of the molecular gas {\it away} from the
nucleus \citep{mcg01,her03}.

In 2001, an upgrade of the 23~GHz system at the VLA enabled us to make
the first interferometric map of the Galactic center in NH\3(6,6)
($\nu=25.056025$~GHz).  With an energy above ground of 412~K,
NH\3(6,6) traces significantly warmer gas than our earlier NH\3 data.
An identical setup to our previous NH\3 observations was used.  After
application of a Gaussian taper to the $(u,v)$ data, the resulting
synthesized beam is $16''\times14''$ \citep{her02}.  A velocity
integrated map of NH\3(6,6) emission is shown in Figure \ref{fig:f1}d.
Contour levels are the same as in panels a--c, and correspond to steps
of $4\sigma_{66}$, where $\sigma_{66}=0.33$~Jy~Beam\1~\kms.  While
some features such as the western streamer are also observed in lower
NH\3 rotation inversion transitions, the velocity integrated image is
dominated by a cloud of hot molecular gas located less than 2~pc in
projected distance from Sgr~A*.  This cloud is apparently located
interior to the CND, and fills the molecular ``hole'' in the NH\3(3,3)
map \citep{her02}.

\section{Derived Parameters \label{derparm}}

\subsection{Opacity and Intrinsic Line Width}

Hyperfine splitting of the NH\3 rotation inversion transitions enables
direct calculation of gas opacity \citep{ho83}.  Each NH\3 rotation
inversion transition is split into five hyperfine components by the
interaction between the electric quadrupole moment of the nitrogen
nucleus and the electric field of the electrons (see \citet{tow75} and
\citet{ho83} for a detailed description of the NH\3 rotation inversion
transitions).  Although spin-spin interactions further split each of
these components, the resulting lines are closely spaced in frequency
and are only resolvable in cold, quiescent cores where line widths are
less than 1--2~\kms.  Therefore, the spectral profile for a typical
molecular cloud with a line width of a few kilometers per second will
consist of a main line and two symmetric pairs of satellite hyperfine
lines, which are located 10--30~km~s$^{-1}$ from the main line.  At
the Galactic center, however, molecular clouds have line widths in
excess of 10~\kms, and even the five electric quadrupole components
are blended.  The blending of these components produces a correlation
between measured line widths of the gas and calculated main line
opacities \citep{mcg02}.

In \citet{mcg02}, we presented a method to disentangle the opacity and
intrinsic line width by using the observed line widths ($\dvo$) and
main line flux densities of two rotation inversion transitions.  In
that paper, we used NH\3(1,1) and (3,3) because they had the highest
signal-to-noise (S/N) (see Figure \ref{fig:f1}).  However, as a result
of their different excitation requirements, the NH\3(3,3) and (1,1)
lines do not in fact trace the same volume of gas at the Galactic
center (see \S \ref{2t}).  We have recalculated the NH\3(1,1) main
line opacity ($\ta$) and $\dvi$ using the NH\3(1,1) and (2,2) data.
With an energy difference of only 42~K, NH\3(1,1) and (2,2) are more
likely to trace the same material.  NH\3(2,2) is significantly fainter
than NH\3(3,3), and the lower S/N limits somewhat the number of pixels
to which our method can be applied.

In addition to changing the two rotation transitions considered in our
method, we also use a Monte-Carlo simulation with 1000 trials to
determine the significance of the derived parameter value at every
pixel.  Solutions are calculated following the algorithm of
\citet{mcg02} for every pixel with a $>3\sigma$ detection of both the
(1,1) and (2,2) lines and an uncertainty in the observed line width of
$<20$~km~s$^{-1}$.  Uncertainties in the measured flux densities are
equal to the rms noise in the map, and uncertainties in the observed
line widths are estimated from the fit of a Gaussian profile to the
spectrum.  Pixels included in the final map must successfully converge
in at least 50\% of the trials.  The exact value of this cutoff does
not appear to be important, and all pixels with solutions $>50\%$ of
the time show similar trends.

For each pixel, the final value for the parameter is taken to be the
mean value from the Monte-Carlo simulations.  The mean values of
$\dvo$ and $\ta$ at each pixel are plotted in Figure \ref{fig:dvtau}a
and c.  No smoothing has been applied to the data, and regions of low
S/N have a ``pixelated'' appearance.  Uncertainties in each parameter
are calculated from the standard deviation of the distribution of
solutions at each pixel.  Figures \ref{fig:dvtau}b and d plot the S/N
of $\dvi$ and $\ta$, respectively, at each pixel.

Table \ref{table:t1} lists the mean and standard deviation of
derived parameters (including $\dvi$ and $\ta$) for the main molecular
clouds in the central 10~pc.  To avoid confusion with mean pixel
values described above, we refer to the cloud-averaged value as the
``characteristic value.''  The characteristic parameter value is
calculated as the average of the parameter at all pixels where a value
was determined from the Monte-Carlo simulations with $>3\sigma$
significance.  For the northern ridge and western streamer, some
parameters had no determinations with $>3\sigma$ significance.  In
these cases, all pixels were used in the calculation of the
characteristic value and standard deviation.  Inclusion of all pixels
will tend to lower the characteristic value because smaller values
will have a lower significance for a given absolute uncertainty.  The
characteristic value quoted in Table \ref{table:t1} is intended to
provide a general value that can be easily compared between different
clouds.  However, it is important to note that many parameters show
significant variations, such as gradients and local maxima/minima,
within a single cloud.  The amount of significant variation of each
parameter within a cloud is characterized by the standard deviation,
which is also listed in Table \ref{table:t1}.

Both the 50 km~s$^{-1}$ GMC ($\Delta\alpha=120''$,
$\Delta\delta=60''$) and the central region of the southern streamer
($30''$, $-90''$) have NH\3(1,1) main line opacities significantly
greater than one (see Figure \ref{fig:dvtau}a).  {The characteristic
opacity that we derive for the southern streamer is very similar to
the values measured by \citet{coi99}.}  The western streamer and
northern ridge have significantly lower opacities with typical values
of $0.3-0.4$.  Hatched regions in Figure \ref{fig:dvtau}b denote
pixels for which only a lower limit for the opacity could be
calculated.  Although some of these values are the result of poor
signal-to-noise, others may reflect real regions of high opacity.
Many of these pixels have main line opacities near five, which
corresponds to the maximum value of opacity considered in our solution
method \citep{mcg02}.

Intrinsic line widths in the central 10~pc tend to have values between
10 and 20~km~s$^{-1}$ (see Figure \ref{fig:dvtau}c).  Significant
increases in intrinsic line widths are observable near many cloud
edges.  However, the apparent random distribution of these features
throughout the central 10~pc makes it difficult to determine the exact
way in which they formed.  In general, these gradients may reflect
dissipative effects at cloud edges possibly due to external pressures
or tidal effects.  For those pixels where only a lower limit on
opacity can be calculated, the effect of blending on observed line
width cannot be determined.  The NH\3(2,2) observed line width is less
affected by blending of the hyperfine lines than $\dvoa$, and it is
therefore used as the upper limit to the intrinsic line width for
these pixels \citep{her.phd}.

Using the derived values of $\ta$ and $\dvi$ along with the measured
main line fluxes, we can calculate the (2,2)-to-(1,1) rotational
temperature, excitation temperature, and column density, and
ultimately estimate cloud masses.  In the derivation of each
parameter, we perform a Monte-Carlo simulation to determine the mean
value and associated uncertainties at each pixel.  Pixels for which
only a limiting value was calculated for $\ta$ and $\dvi$ are not
included in the following calculations.

\subsection{Rotational and Excitation Temperatures \label{temps}}

The ratio of main line emission from two NH\3 rotation inversion
transitions can be used to calculate the rotational temperature of
molecular clouds at the Galactic center.  From \citet{ho83}, the
equation for rotational temperature calculated using NH\3(1,1) and (2,2)
is \bq T_{R21}=\frac{-41.5~\rm{K}}{{\rm
ln}\left(\frac{-0.282}{\tau_m(1,1)}{\rm ln}\left(1-\frac{\Delta
T_{A_m}(2,2)}{\Delta
T_{A_m}(1,1)}\left(1-e^{-\tau_m(1,1)}\right)\right)\right)}~,
\label{eq:t21} \eq where $\frac{\Delta T_{A_m}(2,2)}{\Delta T_{A_m}(1,1)}$
is the ratio of emission from NH\3(2,2) and (1,1) main hyperfine
lines.  (Because of the similar frequencies of the NH\3 transitions,
$\Delta T_A$ depends linearly on flux density, and the ratio of main
line flux densities can be used in place of the ratio of antenna
temperatures.)  Using $\ta$ and the peak fluxes of the NH\3(1,1) and
(2,2) emission, we calculate NH\3 (2,2)-to-(1,1) rotational
temperatures.  Mean values of $T_{R21}$ from the Monte-Carlo
simulations and the associated S/N estimates are plotted in Figure
\ref{fig:temps}a and b.  In addition, Table \ref{table:t1} lists the
characteristic (2,2)-to-(1,1) rotational temperature for each cloud.

(2,2)-to-(1,1) rotational temperatures in the central 10~pc are
generally 25~K with uncertainties on the order of 10\%.  The
relatively small uncertainties are the result of a weak dependence on
$\ta$ for $0<\ta<1$ and (2,2)-to-(1,1) line ratios less than one
\citep{ho83}.  The derived rotational temperature depends primarily on
the (2,2)-to-(1,1) line ratio, which is determined with a much higher
S/N than $\ta$.  The coldest gas is found in the 50 km~s\1 GMC
($120''$, $60''$), where the characteristic rotational temperature is
$21$~K.  The western streamer ($-80''$, $-40''$ and $-50''$, $40''$)
has the warmest gas, with a characteristic (2,2)-to-(1,1) rotational
temperature of $31$~K.

The measured rotational temperature is a lower limit on the true
kinetic temperature of the gas \citep{mar82}.  Table \ref{table:t1}
lists the kinetic temperatures calculated from $\langle
T_{R21}\rangle$ using the relations of \citet{dan88}, which assume a
H\2 density of $10^5$~cm$^{-3}$.  (The degree to which $T_{R21}$
underestimates $T_K$ decreases as the H\2 density increases, until
$T_{R21}\approx T_K$ for $n_{H_2}\simgt10^8$~cm$^{-3}$
\citep{hut93b}.) The resulting range of kinetic temperatures for
clouds in the central 10~pc is roughly 20--50~K.  It is important to
note that NH\3 (2,2)-to-(1,1) rotational temperatures are not
sensitive to gas with kinetic temperatures much greater than 50~K
\citep[see][]{dan88}.  Therefore, the true range of temperatures at
the Galactic center is much larger than indicated by the
(2,2)-to-(1,1) rotational temperatures in Figure \ref{fig:temps}a.
This effect will be discussed in more detail in \S \ref{2t}.

From standard radiative transfer, the excitation temperature is
related to the measured antenna temperature of the line emission,
$\Delta T_A$, by \bq \Delta T_A=\eta~\Phi~(T_{ex}-T_{bg})(1-e^{-\tau})
\eq
\noindent where $\eta$ is the telescope efficiency (0.4 for the VLA at
23~GHz), $\Phi$ is the filling factor, $T_{ex}$ is the excitation
temperature of the gas, and $T_{bg}$ is the temperature of background
radiation incident on the cloud (assumed to be the 2.7~K background of
the Cosmic Microwave Background (CMB)).  Solving for excitation temperature,
we have \bq T_{ex}=2.7~{\rm K}+\frac{\Delta T_A}{\eta \Phi
(1-e^{-\tau})}~.  \eq
\noindent 
Assuming the filling factor is equal to one (probably not a realistic
assumption, see below), we have calculated the excitation temperature
for gas in the central 10~pc.  The resulting maps of mean excitation
temperature and the S/N of these values determined from Monte-Carlo
simulations are shown in Figure \ref{fig:temps}c and d.  Overall, the
excitation temperature has a smooth distribution (especially in
regions with high S/N), with values ranging between 4 and 8~K.  The
characteristic excitation temperatures that we derive are consistent
with the peak values reported by \citet{coi99,coi00}.  Our results are
also consistent with excitation temperatures of $\sim5$~K derived from
NH\3 observations within a few degrees of the Galactic center
\citep{hut93b}.

Throughout the central 10~pc, calculated excitation temperatures are
much lower than the corresponding (2,2)-to-(1,1) rotational
temperatures.  There are two possible explanations for this
discrepancy.  First, low densities could lead to sub-thermal
excitation due to lack of collisions.  From detailed balance of
emission and absorption, the equation for $T_{ex}$ is \bq
e^{-h\nu/k_bT_{ex}}=\frac{\frac{n(H_2)}{n_{crit}}e^{-h\nu/k_bT_K} +
u}{\frac{n(H_2)}{n_{crit}}+u+1}~, \eq where $k_b$ is the Boltzmann
constant and $u=1/(e^{h\nu/k_bT_{rad}} -1)$ \citep[see e.g.][]{stu85}.
The critical density is defined as the ratio of the Einstein
coefficient for spontaneous emission to the collision coefficient.
Using the values for $A_{10}$ from \citet{ho83} and collision
coefficients from \citet{dan88}, we calculate a critical density of
$2\times10^3$~cm$^{-3}$ for every rotation inversion transition that
we have observed.  Figure \ref{fig:textk} plots the excitation
temperature as a function of H\2 volume density for selected kinetic
temperatures.  Despite the low critical density, excitation
temperatures do not increase significantly until
$n(H_2)\approx10^5$~cm$^{-3}$, and the gas is not thermalized
($T_{ex}\approx T_K$) until densities approach $10^6$~cm$^{-3}$.  The
large differences between our measured excitation and kinetic
temperatures imply an H\2 density of $\simlt10^4$~cm$^{-3}$, which is
significantly lower than densities measured for the GMCs and CND.
Sub-thermal excitation of this sort would imply even larger kinetic
temperatures for molecular gas in the central 10~pc.

In \S \ref{un-phys}, we will argue that at least the hottest gas
components in the Galactic center must be nearly thermally excited
with $n$(H\2)$\simgt10^5$~cm$^{-3}$.  A more likely cause of the low
measured excitation temperatures is small filling factors.  If the
clouds fill only a small fraction of the beam, then the observed
excitation temperature will be lowered by the ratio of the cloud area
to the area of the beam.  If the molecular gas in the central 10~pc is
thermalized, then the measured difference between $T_{ex}$ and
$T_{R21}$ implies typical filling factors of $\sim0.2$.

\subsection{Column Densities and Masses}

The NH\3(1,1) column density is related to $\ta$, $T_{ex}$, and $\dvi$
by \citep{ho77} \bq N_{11}=2.8\times10^{13}~{\rm
cm^{-2}}~\tau_m(1,1)~T_{ex}~\left(\frac{\Delta{\rm v}}{1~{\rm
km~s^{-1}}}\right)~.  \eq By correcting for the fraction of molecules
in the (1,1) state, which is characterized by the rotation
temperature, $N_{11}$ can be converted to the total NH\3 column
density, $N_{\rm{NH_3}}$ \citep{tow75}.  Although emission from
non-metastable states has been observed from the nearby Sgr B2
star-forming region \citep{hut93a}, we consider only the lowest 19
metastable states of NH\3 in our calculation of the fractional
population of the NH\3(1,1) transition.
The omission of non-metastable states from our calculation will at
most cause an overestimation of the fractional population of a factor
of two \citep{her.phd}.

Total NH\3 column densities determined from our data are plotted in
Figure \ref{fig:nnh3} and characteristic values for specific clouds
are listed in Table \ref{table:t1}.  Characteristic column densities
for the main features range from $2-9\times10^{15}$~cm$^2$.  However,
the largest column densities ($N_{{\rm NH_3}}>10^{16}$~cm$^2$) are
found primarily in the features associated with the GMCs.  The central
core of the southern streamer ($30''$, $-80''$) has a NH\3 column
density of $1.4(\pm0.3)\times10^{16}$~cm$^{-2}$.  The 50 km~s\1 GMC
also has large NH\3 column densities, with highest values found at the
northeastern edge of our mosaic.  This GMC extends more than an
arcminute past the northeastern edge of our map (see Figure 9 in
\citet{mcg01}), and emission at ($-130''$, $80''$) is associated with
dense regions of the cloud where column densities should be highest.

At the Galactic center, the total mass of a cloud is given by \bq
M=2.4~{\rm M_\odot}~\left(\frac{N_{{\rm NH_3}}}{10^{15}~{\rm
cm}^{-2}}\right)\left(\frac{A}{1~\Box''}\right)\left(\frac{X{\rm
(NH_3)}}{10^{-7}}\right)^{-1}~.  \eq
\noindent where $A$ is the area of the cloud in arcsec$^2$ and
$X$(NH\3) is the abundance of NH\3 relative to H\2.  At present,
$X$(NH\3) is not well-known for Galactic center molecular clouds.
Abundances measured in the Perseus and --3~\kms ~spiral arms and in
dense, star-less cores suggest $X$(NH\3)$=10^{-9}-10^{-8}$
\citep{ben83,batrla84,ser86}.  However, there is much evidence that
NH\3 abundances are enhanced in warm clouds, most likely as the result
of evaporation of NH\3 molecules off of dust grains
\citep{pau83,wal87}.  NH\3 abundances range from $10^{-7}$ to
$10^{-4}$ in warm molecular clouds, with the highest measured
abundances coming from the Sgr B2 complex, one of the largest star
formation regions in the Galaxy \citep[][and references
therein]{hut93a}.  The precise relationship between $X$(NH\3) and
$T_K$ is not straightforward.  Assuming that high temperatures at the
Galactic center will produce enhanced abundances of NH\3, we adopt
$X$(NH\3)=$10^{-7}$ in our calculations.  Resulting estimates of
masses are presented in Table \ref{table:t1}.  The assumption of an
enhanced NH\3 abundance should produce smaller cloud masses than those
calculated by \citet{coi99,coi00}, who use $X$(NH\3)$=10^{-8}$.

A comparison of our derived mass for the edge of the 50 km~s$^{-1}$
GMC to published values in the literature can be used as a consistency
check for our choice of $X$(NH\3).  Total mass estimates for the
50~km~s\1 GMC range between $6\times10^4$ and $4\times10^5$~$M_\odot$
\citep{gus81,arm85,zyl90}.  The virial relation between the mass of a
cloud and its size and line width ($M_V\approx 2\times10^2~{\rm
M_\odot}~\left(\frac{R}{1~{\rm pc}}\right)\left(\frac{\Delta{\rm
v}}{1~{\rm km~s}^{-1}}\right)^2$ for a density profile of $1/r$,
\citet{mac88}) can also be used to calculate the minimum mass of a
gravitationally bound cloud.  Assuming a size of 5~pc and a line width
of 17~km~s\1 (see Table \ref{table:t1}), the virial mass of the
50~km~s\1 GMC is $\sim3\times10^5$~$M_\odot$.  As previously
mentioned, roughly 3/4 of the 50 km~s$^{-1}$ GMC, including the
densest condensations, lies outside our VLA mosaic.  Our derived mass
of $5\times10^4$~$M_\odot$ for the fraction of this cloud within our
mosaic is therefore consistent with what we expect, and we conclude
that $X$(NH\3)$=10^{-7}$ is a reasonable assumption for molecular
clouds near Sgr~A*.

The southern streamer is the second-most massive feature in the
central 10~pc.  We estimate a total mass of $8\times10^4$~$M_\odot$
for the southern streamer.  Using $X$(NH\3)$=10^{-8}$, \citet{coi99}
calculate a mass of $6\times10^4$~$M_\odot$ for gas from the ``central
cloud'' of the southern streamer, which is roughly coincident with the
core of bright NH\3 emission centered at ($+30''$, $-80''$) in our
maps.  If the \citet{coi99} result is re-calculated using our assumed
value of X(NH\3) then the associated mass would be
$6\times10^3$~$M_\odot$, more than an order of magnitude smaller than
our value.  The large difference between these results reflects
different definitions for the size and extent of the southern
streamer.  First, \citet{coi99} assume a size of only 300~arcsec$^2$
for the central cloud, while it appears to be closer to
2000~arcsec$^2$ in our velocity integrated maps.  In addition, the
emission associated with the southern streamer at ($0''$, $-125''$) is
associated with the northern tip of the ``southern cloud'' in
\citet{coi99} and not included in the above mass estimate.  It should
also be noted, that the measured NH\3 column for the central cloud is
consistent between the two papers (for the same assumption of
X(NH\3)).  Considering the different assumptions outlined above, the
reported masses for the southern streamer are consistent.  However, we
believe that our result is a more complete measure of the mass of the
southern streamer from $50''$ to $150''$ south of Sgr A*.

We estimate masses of $1\times10^4$~$M_\odot$ and
$2\times10^4$~$M_\odot$ for SE1 and SE2, respectively.  Together,
these clouds have a mass of a few$\times10^4$~$M_\odot$, and
constitute the majority of the dense ``molecular ridge'' that connects
the 20 km~s$^{-1}$ GMC in the south to the 50 km~s$^{-1}$ GMC in the
northeast \citep{coi00}.  Both the western streamer and northern ridge
have masses roughly an order of magnitude smaller than the southern
streamer and two orders of magnitude smaller than estimated total
masses of the 20 and 50~km~s\1 GMCs \citep{zyl90}.  As we argue in the
following section, we believe these smaller clouds have been swept up
by the expansion of Sgr~A~East.

\section[The Impact of Sgr~A~East]{The Impact of Sgr~A~East on Nearby Molecular Clouds \label{sgeast}}

Maps of radio continuum emission from the Galactic center show a large
shell of synchrotron emission with a radius of $\sim1.6'$ (4~pc)
centered roughly $1'$ to the east of Sgr~A* \citep{ped89}.  This shell
is called Sgr~A~East and is thought to lie within a few parsecs of the
Galactic center (see \S \ref{cartoon}).  To the northeast, the edge of
Sgr~A~East lies precisely along the western edge of the 50 km~s$^{-1}$
GMC, and it is generally agreed that Sgr~A~East is interacting with
the GMC to some extent.  Previous work by \citet{mez89} interpreted
this interaction as indicating that much of the material in the 50
km~s\1 GMC was cleared out of the central parsecs by the expansion of
Sgr A East.  The resulting estimate of the energy of Sgr~A~East is
$\simgt 4\times10^{52}$~ergs, more than an order of magnitude larger
than typical supernova remnants \citep{mez89}.  This result has
sparked many discussions on the possible origin of Sgr~A~East,
including multiple correlated supernovae, expansion within a bubble
\citep{mez89}, or tidal disruption of a star by the supermassive black
hole \citep{khokhlov96}.

The assumption that a significant amount of material in the 50
km~s$^{-1}$ GMC was swept up by the expansion of Sgr~A~East was based
primarily on the apparent concave morphology of the GMC, with the
northern part of the cloud wrapping around the top of the Sgr~A~East
shell \citep{mez89}.  However, in \citet{mcg01}, we showed that this
``northern ridge'' is kinematically distinct from the 50 km~s$^{-1}$ GMC,
with a characteristic velocity of --10~km~s$^{-1}$.  Using the derived
cloud parameters presented in the previous sections, we no longer need
to rely on morphology as the primary indication of an interaction.  In
the following paragraphs, we revisit the question of the impact of Sgr
A East on molecular gas near Sgr~A*, and we calculate a new estimate
for the energy associated with the progenitor explosion.

The 50 km~s$^{-1}$ GMC, northern ridge, and western streamer lie along
the edge of Sgr~A~East \citep{mcg01}.  However, the physical
conditions differ significantly between these clouds.  The molecular
gas in the 50 km~s$^{-1}$ GMC does not appear to be strongly affected
by the impact of Sgr~A~East.  The characteristic (2,2)-to-(1,1)
rotational temperature of the 50~km~s$^{-1}$ GMC ($\langle
T_{R21}\rangle=21$~K) is the lowest value of any feature in the
central 10~pc.  If the 50 km~s$^{-1}$ GMC is moving outwards with the
expansion of Sgr~A~East, then we should expect to see a velocity
gradient along the cloud.  However, no velocity gradient is observed
in this cloud \citep{mcg01}.

The northern ridge and western streamer are more than a magnitude less
massive than the 50 km~s$^{-1}$ GMC, and thus might be expected to be
more affected by the impact of Sgr~A~East.  The northern ridge is a
linear feature, roughly $90''$ in length, that lies along the northern
edge of Sgr~A~East.  Compared to the 50~km~s$^{-1}$ GMC, the northern
ridge has a similar characteristic intrinsic line width
($\langle\dvi\rangle=17$~km~s$^{-1}$) and higher characteristic
(2,2)-to-(1,1) rotational temperature ($\langle T_{R21}\rangle=28$~K).
There is no velocity gradient larger than the size of the spectral
resolution in our data ($\sim9.8$~km~s$^{-1}$), but this result is not
surprising because the cloud velocity of --10~km~s$^{-1}$ is
consistent with the bulk of any expansion occurring perpendicular to
the LOS.

The strongest evidence for an interaction between Sgr~A~East and
molecular gas comes from the western streamer.  The western streamer
also has a long, filamentary structure, and its curvature matches
closely the western edge of Sgr~A~East.  In \citet{mcg01} we reported
a velocity gradient of 1 km~s\1 arcsec\1 (25 km~s\1 pc\1) covering
150$''$ (6~pc) along the western streamer.  At ($60''$, $-80''$)
emission is at $-70$~km~s$^{-1}$, but the velocity smoothly shifts
towards the red until it reaches $\sim90$~km~s$^{-1}$ at ($-70''$,
$70''$) \citep{mcg01}.  A velocity gradient of this type is consistent
with a ridge of gas highly inclined to the line-of-sight that is being
pushed outwards by expansion of Sgr~A~East.  In this model, the
southern part of the streamer would be located on the front side of
the shell, thereby producing the observed blue-shifted velocities.
Molecular gas in the western streamer has the largest characteristic
intrinsic line width ($20$~km~s$^{-1}$) and highest temperatures in
the central 10~pc.  The characteristic (2,2)-to-(1,1) rotational
temperature of $31$~K corresponds to a kinetic temperature near 46~K.
The western streamer is also the only one of these three clouds with
strong NH\3(6,6) emission, and it is likely that some fraction of the
gas has kinetic temperatures $\simgt100$~K.

Based on the above comparison, it seems unlikely that a significant
amount of the 50 km~s$^{-1}$ GMC has been moved by Sgr~A~East.  However,
there is convincing evidence that the expansion of Sgr~A~East swept up
the material in the western streamer.  Assuming that only the western
streamer was cleared out of the central parsecs by Sgr~A~East, we
calculate a new estimate of the age and energy of the progenitor
explosion in the following paragraphs.

A supernova remnant (SNR) is reckoned to have entered the isothermal
or ``snow-plow'' phase when the time scale for radiative cooling
becomes less than the dynamical time of the SNR.  At this point, the
shell expands with constant radial momentum, and a thin, dense layer
of gas is formed that moves outwards with the shock front.  For a SNR
in the snow-plow phase, the energy of the progenitor supernova,
$E_{SN}$, is related to the current energy associated with the shell,
$E_{sh}$, by \citep{shull80} \bq
\left(\frac{E_{SN}}{E_{sh}}\right)=\left(\frac{R_{sh}}{R_{sg}}\right)^2~,
\label{eq:esn}\eq where $R_{sh}$ is the radius
of the shell and $R_{sg}$ is the radius at the time of shell
generation, $t_{sg}$.  The shell radius at time, $t$, is given by
\citep{shull80} \bq R_{sh}=3.29~{\rm
pc}~\left(\frac{E_{SN}}{10^{51}~{\rm
ergs}}\right)^{\frac{1}{4}}\left(\frac{n_o}{100~{\rm
cm}^{-3}}\right)^{-\frac{1}{2}}\left(\frac{t}{t_{sg}}\right)^{\frac{2}{7}}~,
\label{eq:rsh}
\eq where $n_o$ is the mean initial H density in cm$^{-3}$.  At
$t=t_{sg}$, Equation \ref{eq:rsh} defines the shell generation radius,
$R_{sg}$.  Substituting the equation for $R_{sg}$ into Equation
\ref{eq:esn}, and solving for $E_{SN}$ we find \bq
E_{SN}=2.0\times10^{50}~{\rm ergs}~\left(\frac{E_{sh}}{10^{51}~{\rm
ergs}}\right)^{\frac{2}{3}} \left(\frac{R_{sh}}{1~{\rm
pc}}\right)^{\frac{4}{3}} \left(\frac{n_o}{100~{\rm
cm}^{-3}}\right)^{\frac{2}{3}}~.
\label{eq:esn2}\eq

If the $4\times10^3$~$M_\odot$ of material in the western streamer was
swept out of a region with a radius of 4~pc, then the implied mean
initial density is $n_o\approx600$~cm$^{-3}$, which is typical of
molecular clouds.  The kinetic energy associated with the western
streamer (assuming a velocity of 100~km~s$^{-1}$ \citep{mcg01}) is
$E_{ws}=4\times10^{50}$~ergs.  Taking $E_{ws}$ as the shell energy,
the derived energy for the progenitor supernova explosion is
$E_{SN}=2\times10^{51}$~ergs.  This estimate of $E_{SN}$ is a lower
limit on the energy of the progenitor explosion because the western
streamer covers only a fraction of the total shell.  The shell energy
(and therefore $E_{SN}$) will scale as $4\pi/\Omega_{ws}$, where
$\Omega_{ws}$ corresponds to the solid angle initially covered by the
material in the western streamer.  This solid angle likely differs
significantly from the solid angle that it currently covers.  For
example, as the shock front expanded, it may have wrapped around the
densest regions compressing them into narrow filaments and thereby
decreasing the solid angle of the material in the western streamer.
Based on observations of ionized material surrounding Sgr A East, the
total kinetic energy of the shell has been estimated to be
$\sim2\times10^{51}$~ergs \citep{mez89,gen90}.  This kinetic energy
would imply that $\Omega_{ws}/4\pi\approx0.15$.  The mean initial
density calculated from the western streamer is also a lower limit.
However, the observed distribution of dense molecular gas in the
region indicates that Sgr A East probably expanded into a clumpy and
highly non-homogeneous environment.  The filaments of dense gas
detected in NH\3 emission therefore likely account for the majority of
the gas mass surrounding the remnant, and the estimate of $n_o$ from
the western streamer is probably accurate to within a factor of a
few.  Taking $E_{sh}=2\times10^{51}$~ergs and assuming
$n_o=1000$~cm$^{-3}$, we calculate an upper limit on the energy of the
progenitor supernova of $E_{SN}=9\times10^{51}$~ergs.

Based on the range of energies for the progenitor explosion calculated
from our molecular data, we conclude that Sgr A East is most likely
the result of a single supernova that occurred near the Galactic
center.  The expansion of Sgr A East into molecular gas produced the
western streamer and possibly the northern ridge, but it has had
little effect on the majority of the material in the 50 km~s$^{-1}$
GMC.  In fact, this GMC is likely significantly slowing the expansion
of the supernova remnant to the northeast of Sgr~A*.

The case for a single supernova explosion as the progenitor of Sgr A
East has been gaining momentum in the past few years.  In particular,
the small gas mass and thermal energy of $10^{49}$ ergs inferred from
{\it Chandra} observations of hot gas associated with Sgr A East is
consistent with the ejecta of a single supernova \citep{mae02}.  Based
on the centrally concentrated X-ray emission, \citet{mae02} classify
Sgr~A~East as a ``mixed morphology'' (MM) SNR produced by a Type II
supernova.  This classification is supported by the detection of
multiple 1720~MHz OH masers, which are often found in MM SNRs, along
the edge of the shell \citep{yus99,green97}.

Using the estimated energy of the progenitor supernova, we can
calculate the age of Sgr A East.  From \citet{shull80}, the time of
shell generation is given by \bq t_{sg}=3600~{\rm
yr}~\left(\frac{E}{10^{51}~{\rm
ergs}}\right)^{\frac{1}{8}}\left(\frac{n_o}{100~{\rm
cm}^{-3}}\right)^{-\frac{3}{4}}~.  \label{eq:tsg} \eq Substituting the
equation for $t_{sg}$ into Equation \ref{eq:rsh}, we determine an age
of $(1-2)\times10^{4}$~yr for Sgr A East.  This age is similar to the
age of $7500$~yr derived by \citet{mez89} for a model in which the
supernova occurred within a low-density bubble inside the dense 50
km~s$^{-1}$ GMC.  However, the lack of evidence for a strong impact on
the 50~km~s$^{-1}$ GMC leads us to favor a scenario in which the
supernova occurred outside the GMC.  Our derived age of $\sim10^4$~yr
agrees well with other independent measurements of the age of
Sgr~A~East.  Based on the expected differential shearing of a shell
near the Galactic center, \citet{uchida98} independently derive an age
of a few $10^4$ year for Sgr~A~East.  An age of $10^4$~yr is also
consistent with the X-ray properties of the shell \citep{mae02}.

As noted elsewhere \citep[see e.g.][]{ser92,morris96,mae02} the
inferred age of the supernova is 1--2 orders of magnitude smaller than
the age of the H{\sc ii} regions that lie along the western edge of
the 50 km~s$^{-1}$ GMC.  Therefore, these H{\sc ii} regions could not
have been triggered by the impact of the Sgr~A~East shock front on the
molecular cloud.  However, the alignment of these star formation
regions along the edge of the 50 km~s$^{-1}$ GMC is noteworthy, and
may be the result of previous interactions between this cloud and
expanding shock fronts that originated near the nucleus.

\section{The Temperature Structure of the Central 10~pc\label{trots}}

\subsection[A Two-Temperature Gas]{A Two-Temperature Gas  \label{2t}}

In general, a rotational temperature can be calculated from any two
rotation inversion transitions of NH\3 using the equation \bq
T_{Rul}=\frac{-\Delta E_{ul}}{{\rm
ln}\left[\frac{-C_{ul}}{\tau_{m,l}}{\rm ln}\left[1-\frac{\Delta
T_{A_m}(u,u)}{\Delta
T_{A_m}(l,l)}\left(1-e^{-\tau_{m,l}}\right)\right]\right]}~,
\label{eq:trot} \eq
\noindent 
where $\Delta E_{ul}$ is the energy difference between the upper
($J,K=u$) and lower ($J,K=l$) rotation inversion states and
$C_{ul}=\frac{g_l |\mu_{10}|_l^2 k_l}{g_u |\mu_{10}|_u^2 k_u}$ relates
the statistical weights ($g_i$), magnetic dipole moment
($|\mu_{10}|_i^2$), and fraction of emission in the main line ($k_i$)
of the two transitions.  This equation assumes equal excitation
temperatures, filling factors, and telescope efficiencies for the two
transitions.  Due to slow mixing rates ($\sim10^{-6}$~yr$^{-1}$)
between the ``ortho'' ($K=3n$) and ``para'' ($K\ne3n$) spin states of
NH\3, rotational temperatures are generally calculated using two
transitions from the same spin state \citep{ho83}.  Therefore, if the
gas is in LTE, then we can calculate a (6,6)-to-(3,3) rotational
temperature from our VLA data.

NH\3 (6,6)-to-(3,3) line ratios in the central 10~pc are higher than
would be expected based on NH\3 (2,2)-to-(1,1) rotational temperatures
\citep{her02}.  Within 2~pc of Sgr~A*, the (6,6)-to-(3,3) line ratios
exceed theoretical limits, preventing calculation of $T_{R63}$
\citep{her02,her.phd}.  These ``un-physical'' line ratios will be
explained in \S \ref{un-phys}.  Elsewhere in the map, faint NH\3(6,6)
emission makes calculation of $T_{R63}$ difficult (see Figure
\ref{fig:f1}).  In addition, the NH\3(3,3) main line opacity cannot be
reliably calculated from $\ta$ because these two transitions likely
trace gas at different temperatures (see below).  Although $\tc$ is
directly related to the ratio of the main and outer-most satellite
hyperfine lines, the relative population of the satellite lines
compared to the main line is only 0.008 (compared to 0.22 for
NH\3(1,1)), \citep{tow75}), and emission from the satellite hyperfine
lines is generally below the rms noise level of our map.

Due to the above limitations, $T_{R63}$ can only be directly
calculated for the core of the southern streamer and SE1. At ($33''$,
$-75''$), the southern streamer has a measured NH\3 (6,6)-to-(3,3)
line ratio of $0.16\pm0.02$ and a NH\3(3,3) main line opacity of
$\tc=2.5^{+0.4}_{-0.3}$.  The inferred(6,6)-to-(3,3) rotational
temperature of $T_{R63}=80\pm5$~K corresponds to a kinetic temperature
of $100^{+10}_{-~9}$~K for gas with a H\2 density of $10^5$~cm$^{-3}$
\citep{dan88}.  In SE1 ($73''$, $-87''$), the NH\3(3,3) main line
opacity is $3.6^{+1.7}_{-1.5}$.  The (6,6)-to-(3,3) main line ratio of
$0.27^{+0.6}_{-0.5}$ corresponds to $T_{R63}=88^{+22}_{-14}$~K and a
kinetic temperature of $110^{+45}_{-20}$~K.

Lower limits on the kinetic temperature can be calculated in SE2 and
the 50 km~s$^{-1}$ GMC.  In SE2 ($108''$,$-60''$), the
NH\3(6,6)-to-(3,3) line ratio is $0.39^{+0.05}_{-0.04}$, and an upper
limit of 1.5 is calculated for $\tc$.  These values result in lower
limits for the rotational and kinetic temperatures of
$T_{R63}\simgt120$~K and $T_K\simgt190$~K.  For the 50 km~s$^{-1}$
GMC, NH\3(6,6) is only detected near ($112''$, $48''$).  (Although
significant, the velocity integrated NH\3(6,6) emission at this
position does not exceed 1.32~Jy~beam$^{-1}$~km~s$^{-1}$
($\approx4\sigma_{66}$) and thus is not seen in Figure \ref{fig:f1}d.)
At this position, the (6,6)-to-(3,3) line ratio is $0.16\pm0.05$ and
$\tc\simlt2$.  The inferred lower limits on the derived temperatures
are $T_{R63}\simgt75$~K and $T_K\simgt90$~K.  Due to low S/N and large
line widths, the main line opacity cannot be estimated for the western
streamer.  The typical line ratio of 0.7 in this cloud implies an
upper limit of $T_{R63}\simlt230$~K \citep{her02}.  This upper limit
on the (6,6)-to-(3,3) rotational temperature provides little
constraint on the kinetic temperature of the gas \citep{dan88}.

The kinetic temperatures derived above are significantly higher than
those derived from NH\3 (2,2)-to-(1,1) line ratios.  This discrepancy
is indicative of a gas composed of a mixture of separate cool and hot
components.  If multiple temperature components are contained within
the telescope beam, then emission from different rotation inversion
transitions will not trace the same material, and the resultant
measurement of kinetic temperature will depend strongly on the choice
of rotation inversion transitions used in the calculation.  Because
the fractional population of the (J,K) state is a function of kinetic
temperature, each NH\3(J,K) transition is most sensitive to gas at a
particular temperature corresponding to the peak fractional population
of that transition.  The peak fractional population of NH\3(1,1) and
(2,2) occurs around 20--30~K \citep[see][]{her.phd}, making these
transitions excellent tracers of cool molecular material.  At kinetic
temperatures greater than 50~K, however, $T_{R21}$ becomes insensitive
to changes in $T_K$. All kinetic temperatures $\simgt80$~K have
(2,2)-to-(1,1) rotational temperatures between 40 and 60~K
\citep{wal83,ho83,dan88}.  Higher transitions such as NH\3(3,3) and
(6,6) have a small fractional population in cool clouds.  (For
example, only $\sim1$~\% of the NH\3 molecules are in the NH\3(6,6)
state for gas at 50~K.)  The fractional population rapidly increases
towards higher temperatures, making it easier to detect the emission.
In addition, $T_{R63}$ is much more sensitive to variations in kinetic
temperatures between 50 and a few hundred Kelvin.  While an increase
in kinetic temperature from 100 to 200~K changes $T_{R21}$ by only a
few degrees, $T_{R63}$ increases from 80 to 125~K.

Extensive single-dish observations of NH\3 rotation inversion
transitions by \citet{hut93b} show that molecular clouds in the
central $4^\circ\times1^\circ$ of the Galaxy are best characterized by
a two-temperature gas distribution on the scale of 3~pc (corresponding
to the size of the NRAO 40-m primary beam).  In general, 75\% of the
total column of NH\3 comes from a cool gas at 25~K, while the
remaining 25\% of the material is contained in a hot component with a
temperature of roughly 200~K \citep{hut93b}. The uncertainty in these
temperatures is roughly 30\%, but the cold temperature matches well
the observed dust temperature in the region.

Based on the temperatures calculated in this paper, it appears that a
similar cloud structure is valid in the central 10~pc on a spatial
scales of $\sim0.5$~pc.  In Figure \ref{fig:trtk}a, we plot the
inferred kinetic temperature as a function of rotational temperature for
different combinations of hot and cold gas.  The hot gas is assumed to
be at 200~K based on the results of \citet{hut93b}.  The cold
component is set at 15~K in order to be consistent with the kinetic
temperatures inferred from our $T_{R21}$ measurements.  Models for the
derived rotational temperatures are plotted for cases in which the hot
component comprises 0, 25, 50, 75, and 100\% of the gas.  

Mean kinetic temperatures derived from our data are overlaid in Figure
\ref{fig:trtk}a.  The mean kinetic temperature derived from $T_{R21}$
is $26\pm2$~K.  The cold component therefore must have a kinetic
temperature lower than $\sim25$~K.  The two measurements of $T_{R63}$
from our data have a weighted average of $101\pm9$~K.  Although the
low measurements from $T_{R21}$ imply a lower temperature for the cold
component than derived by \citet{hut93b}, the data are consistent with
their conclusions that roughly 25\% of the gas is contained in a hot
component.

In order to study the temperature structure in the central parsecs in
more detail, we have compiled published NH\3 rotational temperatures
from a $7.5'\times11'$ ($17.5\times25.5$~pc) region surrounding Sgr~A*
that includes both the 20 and 50 km~s$^{-1}$ GMC.  A summary of the
references for the measurements, including the telescope and
corresponding beam size, is given in Table \ref{table:t2}, and the
specific values for the rotational temperatures are listed in Table
\ref{table:t3}.  Only one pointing from the large-scale survey by
\citet{hut93b} is contained within this region.

Figure \ref{fig:beams}, overlays the position of each observation on a
1.2~mm continuum image showing bright dust emission from the GMCs and
free-free emission from ionized arcs of Sgr A West \citep{zyl98}.  For
single-dish measurements (Figure \ref{fig:beams}a), the size of the
circle corresponds to the FWHM telescope beam.  Single-dish
observations have focused primarily on the large GMCs, and pointings
tend to be arranged parallel to the Galactic plane.  The FWHM of the
primary beam is plotted for each interferometric observation in Figure
\ref{fig:beams}b.  Both \citet{coi99,coi00} and the data presented in
this paper include a pointing centered on Sgr~A*.  For the
interferometric data, independent temperature measurements may be made
once per synthesized beam.  However, \citet{oku89} and
\citet{coi99,coi00} present calculated temperatures at selected
positions only.  We plot the synthesized beam for the NMA at the
positions of rotational temperature measurements in \citet{oku89}.  In
\citet{coi99,coi00}, characteristic rotational temperatures are given
for the main condensations and we report their estimated positions in
Table \ref{table:t3}.

Using the relation of \citet{dan88}, kinetic temperatures have been
calculated for all measurements of $T_{R21}$, $T_{R42}$, $T_{R54}$,
and $T_{R63}$.  ($T_{R41}$ is not included in \citet{dan88}, and it is
therefore not converted to a kinetic temperature.)  In Figure
\ref{fig:radial}, we plot kinetic temperatures as a function of
distance from Sgr~A*.  The uncertainty in position due to the size of
the beam (see Table \ref{table:t2}) is not plotted.  We also calculate
a radial profile for the kinetic temperatures derived from our
$T_{R21}$ measurements.  The best fit radial profile
$[T_K=31(\pm2)-0.06(\pm0.02)r]$ is overlaid as the grey line in Figure
\ref{fig:radial}a.  The surrounding light grey box shows the
approximate range of derived kinetic temperatures from pixels with a
$\ge3\sigma$ determination of $T_{R21}$.

In general, kinetic temperatures in Figures \ref{fig:radial}a, b, and
d are consistent with a relatively constant kinetic temperature within
$500''$ of Sgr~A*.  Kinetic temperatures derived from our $T_{R21}$
measurements show a slight increase as gas nears Sgr A*.
Unfortunately, there are no measurements of $T_{R63}$, which would be
more sensitive to large increases in the kinetic temperature, within
$100''$ of the nucleus.  Kinetic temperatures derived from $T_{R63}$
for distances between $100''$ and $150''$ are consistent with measured
kinetic temperatures at larger radii.  {\it Any significant heating of
the molecular gas is therefore confined to a region less than
$\sim4$~pc from Sgr~A*.}

In Figure \ref{fig:trtk}b, the average kinetic temperatures calculated
from $T_{R21}$, $T_{R42}$, and $T_{R63}$ measurements in the central
20~pc are plotted as filled circles overlaid on the same models for a
two-temperature gas as Figure \ref{fig:trtk}a.  These temperatures are
calculated as the average of the measurements in Figure
\ref{fig:radial} weighted by the associated uncertainties (see Table
\ref{table:t3}), and are equal to $26\pm6$~K, $55\pm5$~K, and
$100\pm8$~K, for kinetic temperatures derived from $T_{R21}$,
$T_{R42}$, and $T_{R63}$, respectively.  (Values of $T_{R21}$
reported by \citet{gus81} differ significantly from other published
rotational temperatures (see Table \ref{table:t3}) and imply
extraordinarily high kinetic temperatures.  Omission of these data
from our analysis results in $\langle T_{R21}\rangle=24\pm2$~K and
does not affect the conclusions below.)  The mean kinetic
temperatures are consistent with the kinetic temperatures measured
from our VLA data alone.  Based on the consistency between our VLA
measurements and these previously published results, {\it we conclude
that molecular clouds within 10--20~pc of Sgr~A* have a
two-temperature structure on size scales of $\simlt0.5$~pc in which
roughly one quarter of the gas is contained in a hot component.}

Only one measurement of $T_{R54}$ has been made near the 20 or 50
km~s$^{-1}$ GMC \citep{hut93b}.  The inferred kinetic temperature from
this measurement is marked by a triangle in Figure \ref{fig:trtk}b.
The corresponding kinetic temperature derived from $T_{R21}$ at the
same position is also plotted.  Both kinetic temperature measurements
are significantly larger than expected for our two-temperature model.
However, there is a large scatter in $T_{R21}$ measurements in this
region, and it is not surprising that this single measurement set does
not follow the same trend as the weighted average result.

\subsection[The Highest Line Ratios]{Interpreting the Highest Line Ratios at the Galactic Center \label{un-phys}}

Rotational temperatures calculated using Equation \ref{eq:trot} will
give a reasonable value only if \bq 0<\frac{-C_{ul}}{\tau_{m,l}}{\rm
ln}\left[1-R_{m,ul}\left(1-e^{-\tau_{m,l}}\right)\right]<1~, \eq where
$R_{m,ul}=\frac{\Delta T_{A_m}(u,u)}{\Delta T_{A_m}(l,l)}$ is the
observed main line ratio of the upper and lower transitions.  This
constraint translates to an upper limit on the main line ratio of \bq
R_{m,ul}<\frac{1-e^{-\tau_{m,l}/C_{ul}}}{1-e^{-\tau_{m,l}}}~,
\label{eq:rmax} \eq which has a maximum value of $1/C_{ul}$
when $\tau_{m,l}\ll1$.  Thus $1/C_{ul}$ is a firm upper limit on the
observed line ratio, regardless of the opacity or kinetic temperature
of the gas.  We refer to line ratios that exceed this limit as
``un-physical'' because they are not theoretically possible for a
single cloud in LTE.

It is important to note that the two-temperature model described in
the previous section cannot account for un-physical line ratios.  In
that model, gas components at two different temperatures, but each in
LTE, are contained within the telescope beam.  Because the observed
line ratio is simply the weighted average of the line ratios from the
two gas components, it will never exceed the line ratio associated
with the hot cloud component.

Un-physical (6,6)-to-(3,3) line ratios (where $R_{m,63}>2.3$) are
found throughout the central 2~pc.  Figure \ref{fig:pos} shows the
positions of 12 spectra overlaid on velocity integrated NH\3(6,6)
emission.  Figure \ref{fig:spec} plots the corresponding NH\3(3,3)
(grey) and NH\3(6,6) (black) spectra.  The spectra have been Hanning
smoothed, and have an rms noise in one channel of 7.6 and 9.5~mJy for
NH\3(3,3) and (6,6), respectively.  Spectrum A is coincident with Sgr
A*.  In this section, we focus on the spectra associated with
un-physical line ratios.  Gas kinematics will be discussed in the
following section.

Un-physical NH\3 (6,6)-to-(3,3) line ratios occur near --20~\kms ~in
spectra A and B, and +70~\kms ~in spectra D and E.  These line ratios
do not appear to result from an enhancement in NH\3(6,6) emission.
Instead, they occur at velocities where there is no NH\3(3,3)
emission.  This complete lack of NH\3(3,3) emission implies that
un-physical line ratios are likely the result of an absorption process
along the LOS.  The most straightforward scenario for the origin of
un-physical line ratios is one in which emission from a hot cloud with
a high (but physical) (6-6)-to-(3,3) line ratio, passes through cold
molecular material along the line of sight.  The lower-energy NH\3
(3,3) emission will be significantly absorbed by the cold cloud, while
NH\3(6,6) will be relatively unaffected. This model is drawn
schematically in Figure \ref{fig:6633model} and explored in more
detail in the following paragraphs.

For simplicity, we assume that there is no radiative coupling between
the two clouds in our model.  From basic radiative transfer, the
brightness temperature of the first cloud is expressed as \bq
T_{b,1}=T_{ex,1}(1-e^{-\tau_1})+T_{bg}e^{-\tau_1}~. \eq
\noindent For the cold cloud we assume that the background temperature
is equal to the temperature of the CMB. If Cloud 2 is near Cloud 1,
then emission from Cloud 1 will dominate the background, and the
brightness temperature of the second cloud is \bq
T_{b,2}=T_{ex,2}(1-e^{-\tau_2})+T_{b,1}e^{-\tau_2}~.  \eq

The antenna temperatures ``on'' and ``off'' source are related to the
brightness temperature of Cloud 2 by
\begin{eqnarray}
T_{A,{\rm on}}&=&\eta\left[ \Phi T_{b,2}+(1-\Phi)T_{{\rm CMB}}\right] \\
T_{A,{\rm off}}&=&\eta T_{{\rm CMB}}~,
\end{eqnarray}
\noindent where $\eta$ is the telescope efficiency and $\Phi$ is the
filling factor of the source in the telescope beam.  The difference
between these two antenna temperatures will correspond to the measured
antenna temperature due to the source ($\Delta T_A$).  In practice,
$\Delta T_A$ must be corrected for the fraction of emission in the
main hyperfine component ($k$).  For NH\3(3,3) and (6,6), however, the
population of satellite hyperfine lines relative to the main component
is very small and can be ignored ($k_6/k_3=1.08$).  Assuming equal
efficiencies and filling factors for NH\3(3,3) and (6,6), the observed
main line ratio is therefore \bq R_{m,63}\approx\frac{\Delta
T_A(6,6)}{\Delta T_A(3,3)}=\frac{T_{b,2}(6,6)-T_{{\rm
CMB}}}{T_{b,2}(3,3)-T_{{\rm CMB}}}~.  \eq

In order to produce un-physical line ratios, Cloud 2 must have
$\tau_2(3,3)\gg1$ and $\tau_2(6,6)\ll1$.  This can be accomplished
most efficiently for a cold cloud with $T_C\approx 30$~K.  Kinetic
temperatures $<30$~K have only minimal populations in both
transitions, and although the relative population of NH\3(3,3)
compared to NH\3(6,6) is higher at temperatures lower than 30~K, the
small population of these transitions requires a large H\2 column
density.  Clouds with temperatures greater than 40~K will begin to
have a significant population of NH\3(6,6) and both lines will be
affected by absorption in the foreground cloud.

Assuming $\tau_2(3,3)\gg1$, the brightness temperature of the second
cloud will be approximately equal to $T_{ex,2}(3,3)$.  For NH\3(6,6),
however, $\tau_2(6,6)\ll1$ implies that $T_{b,2}(6,6)\approx
T_{b,1}(6,6)$.  Therefore, the observed line ratio can be
approximately written as \bq R_{m,63}\approx\frac{T_{b,1}(6,6)-T_{{\rm
CMB}}}{T_{ex,2}(3,3)-T_{{\rm CMB}}}~.  \eq Un-physical line ratios
with $R_{m,63}\gg1$ imply that $T_{b,1}(6,6)\gg T_{ex,2}(3,3)$.
Assuming that the NH\3(6,6) excitation temperature in Cloud 1 exceeds
$T_{{\rm CMB}}$, then Equation 16 implies that $T_{b,1}(6,6)\le
T_{ex,1}(6,6)$. Therefore, un-physical line ratios will generally
imply that $T_{ex,1}(6,6)\gg T_{ex,2}(3,3)$.  As seen in \S
\ref{temps}, such large differences in excitation temperatures imply
that the H\2 density must exceed $10^5$~cm$^{-3}$ in these clouds (see
Figure \ref{fig:textk}).

It should be noted that un-physical (6,6)-to-(3,3) line ratios were
not reported in the survey of Galactic center molecular clouds by
\citet{arm85}.  However, these data do not include any molecular gas
within 2~pc of Sgr~A* (see Figure \ref{fig:radial}d).  In addition,
the single-dish measurements have a resolution of only $84''$, and
thus sample the average temperature structure on scales of $\sim3$~pc.
Clouds with high or un-physical line ratios are likely to be small and
therefore fill only a fraction of the telescope beam.  Cool gas,
however, is prevalent throughout the entire region.  The inclusion of
significant amounts of cool gas within the telescope beam will lower
the observed line ratios.  Because our VLA data have a resolution of
$16''\times14''$, they are more sensitive to the temperature structure
of molecular clouds on small scales.  However, evidence for beam
dilution of high line ratios is still detected at velocities near
20~km~s$^{-1}$. Although some (6,6)-to-(3,3) line ratios greater than
one occur near 20~km~s$^{-1}$ (see Spectra C, D, and E), no line
ratios exceed $\sim2$.  It is likely that any un-physical line ratios
at these velocities are diluted by emission from cool gas in the
20~km~s$^{-1}$ GMC.  If beam dilution greatly affects observed line
ratios, then the cloud that produces the un-physical line ratios must
be significantly smaller than the synthesized beam of our VLA data.

It is also worth noting that the maximum allowed line ratio can exceed
$1/C_{ul}$ if the cloud is not in LTE.  Equation \ref{eq:trot} assumes
equal telescope efficiencies, filling factors, and excitation
temperatures for the two transitions.  If the values of these
parameters for the upper transition exceed those for the lower
transition, then the corresponding limit on the line ratio will exceed
$1/C_{ul}$.  However, because the energy separation of the inversion
doublets is very small, it seems difficult to imagine a scenario in
which the excitation temperatures of the two transitions differ
significantly.  We prefer the absorption model outlined above because
it provides a simple explanation for un-physical line ratios that is
consistent with observed spectral line profiles without having to
invoke non-LTE conditions.

In summary, we believe that the un-physical line ratios observed to
the east and southeast of Sgr~A* are the result of absorption of
NH\3(3,3) emission from a hot cloud near Sgr~A* by cool material along
the LOS.  Although the current resolution of these data prevents a
determination of the exact parameters of this cloud, the un-physical
line ratios require that at least the hot material must be contained
in a dense component with $n_{H_2}\simgt10^5$~cm$^{-3}$.

\section[Separating Gas Components]{Separating Multiple Gas Components Near Sgr~A*\label{hlrllr}}

Spectra taken from the central 2~pc of the Galaxy are very
complicated, with strong evidence for multiple clouds projected along
the LOS (see spectra A--F in Figure \ref{fig:spec}).  The
(6,6)-to-(3,3) line ratios can be used to separate emission from the
absorbed hot cloud from emission associated with cool, un-absorbed
gas.  Direct comparison of line ratios can be difficult because many
features have infinite NH\3 (6,6)-to-(3,3) line ratios.  We therefore
separate these gas components by using the {\it difference} between
the NH\3(3,3) and (6,6) emission.  Gas with $S_\nu(6,6)>S_\nu(3,3)$ is
termed ``high line ratio'' (HLR) gas, while gas with
$S_\nu(6,6)<S_\nu(3,3)$ is called ``low line ratio'' (LLR) gas.  This
designation works well because gas with $S_\nu(6,6)=S_\nu(3,3)$ (or
$R_{m,63}=1$) corresponds to a rotational temperature of 340~K, or a
kinetic temperature of thousands of Kelvin.  Because most NH\3
molecules will be destroyed at these temperatures, (6,6)-to-(3,3) line
ratios greater than one most likely result from absorption of
NH\3(3,3) emission from a hot cloud by cool intervening material.

The difference in NH\3(3,3) and (6,6) flux is calculated at every
pixel ($x,y,v$) where the emission exceeds three sigma in at least one
of the tracers.  Figure \ref{fig:f12} plots velocity integrated
NH\3(6,6) emission associated with each gas component.  The left-hand
panel of Figure \ref{fig:f12} plots velocity integrated map of
NH\3(6,6) emission associated with the LLR component in contours
overlaid on a velocity integrated map of NH\3(3,3) in grey scale.
There is a very high degree of correspondence between the two maps,
confirming that LLR gas traces clouds observed in lower rotation
inversion transitions.  Spectra F--L in Figure \ref{fig:spec} show
characteristic line profiles for LLR gas.  The right-hand panel of
Figure \ref{fig:f12} overlays a velocity integrated map of NH\3(6,6)
emission associated with HLR gas on a velocity integrated map of
HCN(1--0) from \citet{wri01}.  HLR gas is predominantly located within
2~pc of the Galactic center and interior to the CND.  A tongue of HLR
gas also extends to the east of the nucleus ($60''$, $-30''$) with a
velocity of 25~km~s$^{-1}$.  This HLR gas does not appear to be
associated with the streamer at 60~km~s$^{-1}$ and offset roughly
$50''$ to the northeast of Sgr~A* that was detected in HCN(1--0) by
\citet{ho95}.

Figure \ref{fig:pos} shows the location of a position-velocity cut
that begins in the southern streamer and passes northwards through the
central 2~pc of the Galaxy.  The corresponding position velocity
diagram of $S_\nu(6,6)-S_\nu(3,3)$ is shown in grey scale in Figure
\ref{fig:pv}.  Although $S_\nu(6,6)-S_\nu(3,3)$ ranges between --0.65
and 0.13, the grey scale stretch ranges from --0.25 to 0.25 to
highlight the different components in the data.  Black corresponds to
LLR gas, and white corresponds to HLR gas.  The $0''$ position
corresponds to $\Delta\delta=0''$.

Panels $b-d$ of Figure \ref{fig:pv} compare the difference map to
emission from three molecular tracers.  In Figure \ref{fig:pv}b, we
overlay a position-velocity cut from our NH\3(6,6) data cube.
Contours are in steps of $3\sigma_{66,ch}$, where
$\sigma_{66,ch}=0.014$~Jy~Beam\1 \citep{her02}.  Emission from
NH\3(6,6) is present in both the HLR and LLR gas.  Figure
\ref{fig:pv}c plots a position-velocity cut of NH\3(3,3) with
contours in steps of 3, 6, 12, 24, and $48\sigma_{33,ch}$, where
$\sigma_{33,ch}=0.011$~Jy~Beam\1 \citep{mcg01}.  Unlike NH\3(6,6),
NH\3(3,3) emission is only present in the LLR gas.  The lack of
NH\3(3,3) emission is consistent with all of the HLR gas resulting
from absorption of NH\3(3,3) by cool intervening material along the
LOS.

In Figure \ref{fig:pv}d, we plot a position-velocity diagram of
HCN(1--0) from \citet{wri01}.  This position-velocity diagram
resembles neither the NH\3(6,6) nor (3,3) plot.  From $-150''$ to
$-30''$, only the line wings of the southern streamer are detected.
Self-absorption of HCN(1--0) is detected throughout the central 10~pc
of the Galaxy (see spectra A, B, C, D, and F in \citet{wri01}) and
strongly affects emission from the massive clouds \citep{mcg01}.  The
strong self-absorption results from the low equivalent temperature of
the HCN(1--0) transition (5~K).  The higher equivalent temperatures of
NH\3 (equal to $125$~K for NH\3(3,3)) make the rotation inversion
transitions much less susceptible to absorption.  Although not
precisely coincident, the HCN(1--0) emission appears to be most
closely related to emission from NH\3(3,3) and the LLR gas.

\subsection{The Circumnuclear Disk}

Faint NH\3(3,3) emission associated with the CND is detected at
$-30$~km~s$^{-1}$ at $-20''$ in Figure \ref{fig:pv}c.  The CND is the
dominant feature in the HCN(1--0) data, with emission observed from
$-70$~km~s\1 at $-30''$ to $+100$~km~s\1 at $80''$ (see Figure
\ref{fig:pv}d).  However, the HCN(1--0) emission is not coincident
with the NH\3(3,3) emission.  In particular, HCN(1--0) emission at
$-20''$ shows two peaks at $-20$ and $-40$~km~s$^{-1}$ that bracket
the NH\3(3,3) emission at $-30$~km~s$^{-1}$, indicating that
self-absorption of HCN(1--0) is a significant effect even in the CND.

It is well-established that there is a lack of HCN(1--0) emission in
the southeastern part of the CND \citep{gus87,wri01}.  In Figure
\ref{fig:pv}d, we find that emission from HCN(1--0) and HLR gas do not
overlap at any position.  Therefore, it might be suspected that the
same material that absorbs NH\3(3,3) to produce un-physical
(6,6)-to-(3,3) line ratios also absorbs HCN(1--0) emission from the
eastern side of the CND.  However, a detailed comparison of the HCN
and NH\3 data in the ``missing'' southeastern part of the CND indicate
that this scenario is unlikely.  Both HCN(1--0) and HLR gas are absent
at a position of ($40''$, $-10''$) relative to Sgr~A* (see Figure
\ref{fig:f12}).  At ($25''$, $-20''$) HLR gas overlaps part of the
missing ring, but it has velocities from $30-100$~\kms~ and will not
absorb emission from the CND, which is expected to have velocities
between $\pm20$~\kms (see Figure 12 of \citet{wri01}).  Finally, lower
line ratios near 20~km~s$^{-1}$ indicate that the HLR gas cloud has a
small filling factor compared to the southern streamer.  It is thus
unlikely that the cloud could absorb a significant portion of the CND.
The ``missing'' southeastern part of the CND most likely reflects an
intrinsically clumpy morphology.

\subsection{The Southern Streamer \label{ss}}

The majority of LLR gas in Figure \ref{fig:pv} is associated with the
southern streamer.  This northeastern extension of the 20 km~s\1 GMC
towards the CND has been suggested to be physically connected to the
nucleus \citep{ho91,coi99,mcg01}.  However, this model has been
hindered by a small velocity gradient and only minimal heating.  For a
central mass of $10^7~M_\odot$, a cloud at a distance of 5~pc should
have a velocity gradient of $\sim9$~km~s$^{-1}$~pc$^{-1}$
\citep{coi99}.  However, a velocity gradient of only
2--3~km~s$^{-1}$~pc$^{-1}$ is measured along the 5~pc length of this
cloud \citep{coi99}.  The small observed velocity gradient implies
that the bulk of the acceleration must occur perpendicular to the LOS.
This orientation seems unlikely, however, because the 20 km~s$^{-1}$
GMC is known to be located in front of the Galactic center along the
LOS (see \S \ref{cartoon}).

LLR gas associated with the southern streamer has a typical velocity
of 20--30~km~s$^{-1}$.  Strong NH\3(3,3) emission is associated with
this feature (see Figure \ref{fig:pv}c), and even the outer satellite
hyperfine lines, which are offset by $\pm29$~km~s\1 from the main
line, are detected.  Emission from the southern streamer appears to
continue to the north of the CND, with no evidence for a significant
change in velocity.  LLR gas can also be seen in Spectrum A of Figure
\ref{fig:spec}.  In this spectrum, the LLR gas shows slightly higher
velocities of $\sim50$~km~s$^{-1}$.  This molecular gas at 50
km~s$^{-1}$ is most likely responsible for the stimulated H~92$\alpha$
emission observed towards Sgr~A* by \citet{rob93}.  Assuming that the
LLR gas in Spectrum A is associated with the southern streamer, we
measure an upper limit for the velocity gradient along the streamer of
12~km~s$^{-1}$~arcmin$^{-1}$ (5~km~s$^{-1}$~pc$^{-1}$), which is still
a factor of two smaller than the expected gradient for infalling gas.
Increased line widths near the CND were also reported by
\citep{coi99}.  In NH\3(3,3) we see no significant evidence for line
width broadening associated with the LLR gas.  Based on the
continuation of the streamer to the north of Sgr~A* as well as the
lack of a large velocity gradient or increased line widths, we
conclude that the southern streamer shows no conclusive evidence for
direct interaction with material in the central 2~pc of the Galaxy.
Instead, its apparent approach towards the Galactic center is most
likely the result of projection along the LOS in front of Sgr~A*.

Although we favor the above model for the southern streamer, the
$15''$ resolution of our data prevents us from unequivocally ruling
out an interaction of some type with the nucleus.  While the
kinematics of the southern streamer do not indicate a close
association with the nucleus, it should be noted that a significant
change in the NH\3(3,3) emission occurs at roughly the location of the
southern lobe of the CND.  NH\3(3,3) emission north of the CND is
significantly weaker than the emission to the south.  The coincidence
of this change with the nucleus may indicate that there is some
interaction between this cloud and the nucleus.  

It is also possible that the faint LLR emission to the north of Sgr~A*
is not associated with the southern streamer at all.  LLR emission
from $0-50''$ in Figure \ref{fig:pv}c appears to fill in the
velocities between the two lobes of high velocity emission associated
with the CND.  Continuous emission from $-30''$ to $+80''$ would imply
that the CND extends very close to the nucleus, which is contrary to
the current model of the CND as a ring-like structure.  However,
observations of HCN(1--0) emission from the CND have shown that it is
warped and clumpy, and it differs significantly from a simple
Keplerian ring \citep{gus87}.  It has even been suggested that the
structure is the result of three independent clouds in orbit about the
nucleus \citep{wri01}.  A more precise understanding of the CND is
necessary to rule out an association with the 30~km~s$^{-1}$ LLR gas
to the north of Sgr~A*.

Finally, the faint NH\3(3,3) emission to the north of Sgr~A* may
partially result from ``blooming'' of flux during the MEM
deconvolution.  (Blooming occurs when extra flux in the form of a
faint, extended component is added to the image by MEM in order to
obtain the ``best'' solution.)  Because emission to the north of the
CND has slightly more red-shifted velocities than the bright southern
emission, we believe that it is unlikely that this emission is the
result of blooming.  These questions should be resolved with the
future addition of single-dish data, which will eliminate
deconvolution artifacts and also allow us to use the full $3''$
resolution of the data.

\subsection{The High-Line-Ratio Cloud}

High line ratio gas has a very different kinematic structure from the
southern streamer and CND.  Figure \ref{fig:pv} shows that HLR gas has
a large velocity gradient from $\simgt50$~km~s\1 at $-20''$ to
--20~\kms ~at $\sim0''$ and then back to 0~\kms ~at $+20''$.  Due to
the large range of velocities over which HLR gas is observed, a cool
gas cloud along the LOS cannot produce the majority of the absorption
of NH\3(3,3).  {\it The absorbing material responsible for the
un-physical (6,6)-to-(3,3) line ratios most likely lies in a shielded
outer layer of the same cloud.}  The rotation of the HLR cloud is in
the opposite sense as the CND.  The gradient is in the same direction
as the high velocity cloud discussed in \citet{mcg01}, but the
positions and velocities of the emission from the two features do not
precisely agree.  Improved spatial resolution is necessary to
determine the exact morphology of this cloud.

The large velocity gradient in the HLR gas explains the observed line
widths of of 80--90~km~s\1 in Figure 2 of \citet{her02}.  The true
line width of the gas can be measured at the turning point in the
NH\3(6,6) position-velocity diagram (position of $0''$ in Figure
\ref{fig:pv}b).  We estimate that the intrinsic line width of the
HLR gas is $\sim20$~km~s\1.  This line width is very similar to
intrinsic line widths determined for the other molecular features in
the central 10~pc (see Figure \ref{fig:dvtau}c).

Line ratios at the position where HLR gas crosses the 20 km~s$^{-1}$
GMC favor a model in which the HLR cloud is significantly smaller than
the synthesized beam of these data.  At a position of $-15''$ in
Figure \ref{fig:pv}a, there is a slight decrease in the values of
$S_\nu(6,6)-S_\nu(3,3)$.  Lower (6,6)-to-(3,3) line ratios are
consistent with beam dilution of line ratios at these velocities by
cool emission from the southern streamer.

The ``hook'' profile of the position-velocity diagram for the HLR gas
is characteristic of a cloud that is undergoing both rotation and
either infall or expansion \citep{ket88}.  The position-velocity
profile of the HLR gas is best fit by a linear velocity gradient of
--1.6~km~s\1~arcsec\1 ($-40$~km~s\1~pc\1) and expansion or contraction
at 80~\kms.  Assuming the gradient is due to circular rotation and
that the measured length of the cloud ($\sim1$~pc) is twice the
radius, the observed gradient corresponds to rotation at 20~km~s\1.
This rotation velocity implies a minimum enclosed mass of
$4\times10^4$~$M_\odot$.  If the cloud is assumed to have a circular
orbit around a mass of $10^7$~$M_\odot$, then the observed velocity
gradient would correspond to an inclination of $\sim4^\circ$.  Such a
low inclination is unlikely, although it cannot be ruled out with the
current resolution, and the HLR cloud is probably not in a circular
orbit.

Because emission is blue-shifted from the expected rotation
velocities, we must either be observing emission from the far side of
a contracting cloud or the near side of an expanding shell.  We
currently favor a model in which we are observing the near edge of an
expanding shell.  Assuming the source of the expansion is therefore
behind the observed shell along the line-of-sight, then it naturally
follows that the cool layer which absorbs the NH\3(3,3) would be
located between the hot material on the inner edge of the shell and
the observer.  Although it is possible that the other half of the
shell is truly missing, the lack of symmetry may also provide clues to
the morphology of the shell and the location of the cool absorbing
material.  Improved spatial resolution will be useful in resolving
this question.

\section{Towards a 3-D Model of the Central 10~pc \label{cartoon}}

In the preceding sections, we have investigated the physical
conditions and interactions of dense molecular clouds within 5~pc of
Sgr~A*.  We briefly summarize our main results below:
\begin{itemize}
\item The 50~km~s$^{-1}$ GMC shows no evidence for a strong
interaction with Sgr A East.  Based on the observed effects of Sgr A
East on the western streamer, we find that the energy of Sgr A East is
consistent with a single supernova occurring roughly $10^4$ years ago
(\S \ref{sgeast}).
\item The temperature structure of dense clouds in the central 10~pc
can be characterized by a two-temperature structure on scales of
0.5~pc.  The data are consistent with the structure on scales of 3~pc
proposed by \citep{hut93b}, in which roughly one quarter of the gas is
in a hot component at 200~K.
\item NH\3 (6,6)-to-(3,3) line ratios in the central 2~pc exceed
theoretical limits for gas in LTE (\S \ref{un-phys}).  We conclude
that these line ratios result from absorption of NH\3(3,3) by a cool,
shielded layer of the same cloud.  This result implies that
$n_{H_2}\simgt10^5$~cm$^{-3}$.
\item Three distinct molecular gas components lie within a projected
distance of 2~pc from Sgr A* (\S \ref{hlrllr}).  Both the CND and the
``high-line-ratio gas'' show large velocity gradients, although in
opposite directions.  Emission associated with the southern streamer,
however, shows no significant velocity gradient, and extends to the
north of the nucleus.  
\end{itemize}

It is important to relate the molecular features described in this
paper to other structures in the nuclear region.  Discussions of the
central parsecs of the Galaxy seem to inevitably evolve into
discussions of the location of these structures along the LOS.  In
this concluding section, we present a model for the LOS arrangement of
the features in the central 10~pc.  The model is based on the
molecular line data presented in this paper as well as the most
current results in the published literature.  The model is not to
scale (or completely inclusive), but it is presented to promote
discussion of the interaction between different structures observed at
the Galactic center.

A schematic drawing of the main structures in the central 10~pc of the
Galaxy is presented in Figure \ref{fig:f17}.  The elongated Sgr~A~East
shell fills the center of the image.  Towards the western side of the
shell, a black dot shows the location of Sgr~A*, which is surrounded
by the arcs of Sgr A West (light grey) and the backwards ``C'' shape
of the CND (black).  The cloud of HLR gas discussed in \S \ref{hlrllr}
is also shown in close proximity to Sgr~A*.  Because its exact
morphology is not known, it is shown as a grey circle, slightly offset
to the southeast of Sgr~A*.  The major molecular features described in
this paper are labeled.  Hatched lines extending northwards from the
southern streamer indicate the faint extension of this cloud to the
north of Sgr~A* (see \S \ref{ss}).  A second supernova, G359.92--0.09,
which is impacting Sgr~A~East in the southeast is also shown
\citep{coi00}.  Positions of 1720~MHz OH masers \citep{yus99} are
marked with crosses, and the H{\sc ii} regions observable at 6~cm are
shown as small circles in the 50~km~s\1 GMC.

In Figures \ref{fig:f18} and \ref{fig:f19}, we present schematic
drawings of the proposed location of these structures along the LOS.
Figure \ref{fig:f19} also shows the motions of the clouds inferred
from their Doppler-shifted velocities.  The LOS structure of the arcs
of Sgr A West is based on H~92$\alpha$ emission from \citet{rob93}.
Because the location of the eastern bar relative to Sgr~A* is not well
known, it is not included in the LOS model.  The northern arc is best
described as a tidally disrupted cloud that is passing within 0.13~pc
of Sgr~A*.  The orbit produces high negative velocities of --250
km~s$^{-1}$ at the southern-most end of the arc \citep{rob93,rob96}.
The western arc, however, is coincident with the inner edge of the
circumnuclear disk, and most likely is the ionized inner layer of this
structure \citep{rob93}.  Absorption of starlight and Br$\gamma$
emission from the western arc of Sgr A West favors an orientation of
the CND with the western side in front of the Galactic center along
the LOS \citep{rob93}.  The location of the HLR molecular gas is
schematically drawn interior to the CND, but exact determination of
the morphology and location of this cloud cannot be determined with
the current resolution of our data.

Absorption of 90~cm continuum emission from Sgr~A~East by Sgr A West
places the nucleus in front or just inside Sgr~A~East \citep{ped89}.
The connection of the northern ridge, which lies along the northern
edge of Sgr~A~East, to the CND indicates that Sgr A East must be
within a few parsecs of the nucleus \citep{mcg01}.  More recently,
\citet{mae02} place Sgr~A*, Sgr A West, and the CND just inside the
leading edge of Sgr~A~East.  (For schematic drawings of proposed
relative locations of Sgr A West, Sgr~A*, and Sgr~A~East, see e.g.
\citet{ped89} and \citet{mae02}.)

As discussed in \S \ref{sgeast}, the velocity gradient of the western
streamer indicates that it is highly inclined to the LOS and expanding
outwards with Sgr~A~East.  The northern ridge is placed along the
northern edge of Sgr~A~East.  The cloud is located at the same
distance as the center of Sgr~A~East, with an orientation roughly in
the plane of the sky.  This placement allows for the bulk of the
motion associated with the expansion of Sgr~A~East to occur
perpendicular to the line of sight, consistent with its velocity of
--10~km~s$^{-1}$.

The 50~km~s\1 cloud is located predominately to the east of
Sgr~A~East.  As discussed in \S \ref{hlrllr}, stimulated H~92$\alpha$
emission at 50~km~s$^{-1}$ observed towards Sgr~A* \citep{rob93} is
most likely associated with the southern streamer and not the GMC.
The 50~km~s\1 cloud is not observed to strongly absorb X-ray emission
from the central 10~pc \citep{park04}, which also implies that little
of the material is in front of the Galactic center along the LOS.
This GMC is connected to the 20 km~s\1 cloud along the molecular ridge
\citep{coi00}.  As mentioned in \S \ref{sgeast}, the H{\sc ii} regions
found along the interior of the western edge of this GMC are much too
old to have been caused by the current impact of Sgr A East.

Observations of formaldehyde absorption towards the Galactic center,
indicate that the 20 km~s$^{-1}$ GMC lies in front of the nucleus
along the LOS \citep{gus80}.  This placement has been recently
confirmed by absorption of 2--10~keV X-rays in the region of this GMC
\citep{park04}.  The distance between the 20 km~s$^{-1}$ GMC and the
nucleus is constrained by observations of SNR 359.92--0.09.  This SNR
was first reported by \citet{coi00}, who detected it in 20~cm radio
continuum images from \citet{yus87}.  More recently, an X-ray filament
associated with the southwest edge of this SNR has been detected with
both {\it XMM-Newton} and {\it Chandra} \citep{sakano03,park04}.  The
impact of SNR 359.92--0.09 is responsible for the concave morphology
of Sgr~A~East in the southeast.  Multiple 1720~MHz OH masers with
velocities of $\sim50$~km~s$^{-1}$ have been detected along this
interface region \citep{yus99}.  These masers have been associated
with disrupted gas associated with the 20~km~s$^{-1}$ GMC
\citep{coi00}.  Because SNR 359.92--0.09 is interacting with both
Sgr~A~East and the 20 km~s$^{-1}$ GMC, the GMC must be no more than
$5-10$~pc from Sgr~A*.  Finally, based on the results of \S
\ref{hlrllr}, the southern streamer is depicted as extending
northwards from the 20~km~s\1 cloud, but not interacting with the
nucleus.

Although uncertainties in this model undoubtedly remain, the fact that
such a detailed three dimensional model of the Galactic center can be
proposed based on observational data reflects the great advances in
recent years in our understanding of this region.  In particular, the
data presented in this paper have shed new light on the role of dense
molecular gas in the region around the supermassive black hole.  The
large velocity coverage of our VLA data combined with the relatively
high excitation requirements of the rotation inversion transitions
have resulted in a more complete picture of the location of dense
material within 5~pc of Sgr~A*.  The additional knowledge of
temperatures, line widths, and cloud masses, all of which are directly
obtained from the NH\3 data, have enabled us to more precisely
determine the interactions between these dense molecular clouds and
other features near the nucleus.  At present, a more precise
understanding of the role of the dense molecular structures in the
Galactic center is primarily hindered by the relatively low resolution
($16''\times14''$) of our data.  Future incorporation of measurements
made with large single-dish telescopes should allow us to resolve many
of the outstanding questions regarding dense molecular gas at the
center of the Galaxy.

\acknowledgments The authors would like to thank R. Zylka for the
1.2~mm continuum image, M. Wright for the HCN(1--0) data, and the
referee for a very careful and thorough report.  R. M. H. would also
like to thank J. Greene and P. Sollins for many productive
discussions.  The 6~cm continuum image from \citet{yus87} was obtained
from the ADIL library.

\bibliographystyle{apj}
\bibliography{Herrnstein}

\begin{deluxetable}{l c r r r r r r r}
\singlespace
\tabletypesize{\scriptsize}
\tablewidth{0pt}
\tablecaption{Physical Parameters for Galactic Center Clouds\tablenotemark{a} \label{table:t1}}
\tablehead{ 
\colhead{Parameter} & \colhead{Unit} & \colhead{Southern} & \colhead{Northern} & \colhead{SE1} & \colhead{SE2} & \colhead{Northern} & 
\colhead{50 km~s$^{-1}$} & 
\colhead{Western} \\
\colhead{} & \colhead{} & \colhead{Streamer} & \colhead{Fork} & \colhead{} & \colhead{} & \colhead{Ridge} & \colhead{GMC} & \colhead{Streamer} 
}
\startdata
Extent in R.A. & $''$ & --30, 60 & --70, --30 & 60, 90 & 90, 140 & 30, 100 & 80, 150 & --100, --40 \\
Extent in Dec. & $''$ & --30, -150 & --150, -90 & --30, --140 & --10, --100 & 80, 130 & 0, 125 & --100, 100 \\
$\langle\Delta{\rm v}_{{\rm int}}\rangle$ & km s$^{-1}$ & 15.2& 13.9& 14.6& 16.7& 15.8& 16.9& 20.5\\
$\sigma(\Delta{\rm v}_{{\rm int}})$ & km s$^{-1}$       & 8.1 & 5.4 & 8.1 & 10.0& 4.0 & 6.4& 4.5 \\
$ \langle\tau_m(1,1)\rangle$ &     ...                  & 2.4 & 3.0 & 2.2 & 1.9 & 0.5\tablenotemark{e}& 2.5& 0.4\tablenotemark{e}\\
$\sigma(\tau_m(1,1))$ &            ...                  & 0.7 & 0.2 & 0.7 & 0.3 & 0.4\tablenotemark{e}& 0.5& 0.3\tablenotemark{e}\\
$ \langle T_{R21}\rangle$ & K                           & 26.4& 25.4& 28.9& 29.3& 28.2& 21.4& 30.9\\
$ \sigma(T_{R21})$ & K                                  & 5.1 & 2.8 & 5.2 & 4.7 & 4.5 & 2.9 & 3.9 \\
$ \langle T_{K}\rangle$\tablenotemark{b} & K     & 33~~~& 32~~~& 39~~~& 40~~~& 37~~~& 25~~~& 46~~~\\	             
$ \langle T_{ex}\rangle$ & K                            & 6.2 & 4.3 & 4.9 & 4.7 & 4.9\tablenotemark{e}& 4.4 & 4.1\tablenotemark{e}\\
$ \sigma(T_{ex})$ & K                                   & 1.1 & 0.2 & 0.5 & 0.6 & 1.4\tablenotemark{e}& 0.5 & 0.7\tablenotemark{e}\\
$ \langle N_{(1,1)}\rangle$ & $10^{15}$ cm$^{-2}$       & 3.5 & 3.6 & 1.5 & 1.2 & 0.9 & 2.8 & 1.4\tablenotemark{e}\\
$\sigma(N_{(1,1)})$ & $10^{15}$ cm$^{-2}$               & 1.9 & 1.0 & 1.1 & 0.9 & 0.2 & 1.1 & 0.6\tablenotemark{e}\\
$ \langle N_{{\rm NH_3}}\rangle$ & $10^{15}$ cm$^{-2}$  & 8.7 & 5.0 & 3.0 & 2.9 & 2.1 & 6.1 & 3.7\tablenotemark{e}\\
$ \sigma(N_{{\rm NH_3}})$ & $10^{15}$ cm$^{-2}$         & 4.3 & 4.2 & 2.1 & 2.1 & 0.5 & 2.6 & 1.4\tablenotemark{e}\\
Area\tablenotemark{c} & $10^3\Box''$                                    & 4.0 & 0.3 & 1.5 & 2.2 & 0.4 & 3.3 & 0.4 \\
Mass\tablenotemark{d}& $M_\odot$                       & $8\times10^4$ & $4\times10^3$ & 1$\times10^4$ & $2\times10^4$ & $2\times10^3$ & $5\times10^4$ & $4\times10^3$ \\

\enddata 
\tablenotetext{a}{The mean and standard deviation of each
parameter is calculated using all pixels where a value was determined
from Monte-Carlo simulations with $>3\sigma$ significance.}
\tablenotetext{b}{$\langle T_K\rangle$ is calculated from $\langle T_{R21}\rangle$ using \citet{dan88}, which assumes $n({\rm H}_2)=10^5$~cm$^{-3}$.} 
\tablenotetext{c}{Cloud area is calculated from the number of pixels with $N_{NH_3}$ determinations (see Figure \ref{fig:nnh3}a).}
\tablenotetext{d}{Masses are calculated
assuming $X$(NH\3)$=10^{-7}$.  Because the value of $X$(NH\3) is not
well-determined, masses should be taken as order of magnitude
estimates.}
\tablenotetext{e}{No pixels with $>3\sigma$ significance for
this feature.  All pixels with a determination from the Monte-Carlo
simulation are included in the calculation.}
\end{deluxetable}

\begin{deluxetable}{l c r c c}
\tablecaption{Summary of NH\3 Observations at the Galactic Center \label{table:t2}}
\tablehead{ 
\colhead{Label} & \colhead{Telescope} & \colhead{Beam\tablenotemark{\dagger} ($''$)} & \colhead{Measured Temperatures} & \colhead{Reference}
}
\startdata
G81    & Effelsberg & 40            & $T_{R21}$, $T_{R41}$, $T_{R42}$ & \citet{gus81} \\
O89    & NMA        & $24\times 13$ & $T_{R21}$ & \citet{oku89} \\
A85    & Haystack   & 84            & $T_{R21}$, $T_{R63}$ & \citet{arm85} \\
H93    & NRAO 40-m & 84             & $T_{R21}$, $T_{R54}$ & \citet{hut93b} \\
C99/00 & VLA                        & $14\times9$\tablenotemark{\ddagger}  & $T_{R21}$ & \citet{coi99,coi00} \\
H04    & VLA                        & $15\times13$ & $T_{R21}$ & this paper \\
\enddata
\tablenotetext{\dagger}{For interferometric observations, this  synthesized beam is listed.}
\tablenotetext{\ddagger}{\citet{coi99,coi00} report average values for each condensation, which is typically $30''$ in size.}
\end{deluxetable}  

\begin{deluxetable}{r r c c c c c  c}
\singlespace
\tabletypesize{\small}
\tablewidth{0pt}
\tablecaption{NH\3 Rotational Temperature Measurements for Galactic Center Molecular Clouds \label{table:t3}}
\tablehead{ 
\colhead{$\Delta\alpha$} & \colhead{$\Delta\delta$}  & \colhead{$T_{R21}$} & \colhead{$T_{R41}$} & \colhead{$T_{R42}$} & \colhead{$T_{R54}$} & \colhead{$T_{R63}$} & \colhead{Ref.}\\
\colhead{($''$)} & \colhead{($''$)}  & \colhead{(K)} & \colhead{(K)} & \colhead{(K)} & \colhead{(K)} & \colhead{(K)} 
}
\startdata
$185$   &  $113$    &  ...    & $>47$   & ...    & ...  & ...         & G81 \\
$76$    &  $95$     &  ...    & $>44$   & $>41$  & ...  & ...         & G81 \\
$143$   &  $53$     &  ...    & $>52$   & $>54$  & ...  & ...         & G81 \\
$152$   &  $47$     &  ...    & $>49$   & $>56$  & ...  & ...         & G81 \\
$177$   &  $32$     &  ...    & $>39$   & $>43$  & ...  & ...         & G81 \\
$122$   &  $19$     &  ...    & $>49$   & $>61$  & ...  & ...         & G81 \\
$62$    &  $-155$   &  66\tablenotemark{\dagger }     & 46    & 46   & ...    & ...       & G81 \\
$30$    &  $-207$   &  ...    & ...   & 49   & ...    & ...       & G81 \\
$-3$    &  $-259$   & 	$55^{+15}_{-10}$  & $45^{+15}_{-10}$ & $49^{+10}_{-6}$ & ...& ...  & G81 \\
$-67$   &  $-287$   &  96     & 47    & 43   & ...   & ...       & G81 \\
$-33$   &  $-308$   &  $100^{+51}_{-25}$ & $40^{+4}_{-3}$   & $37^{+6}_{-4}$  & ...& ...  & G81 \\
$-66$   &  $-360$   &  71     & 46    & 43   & ...   & ...        & G81 \\
$-13$   &  $-392$   &  63     & 44    & 46   & ...   & ...        & G81 \\
$-97$   &  $-412$   &  91     & 40    & 37   & ...   & ...        & G81 \\
$39$    &  $-424$   &  51     & 41    & 46   & ...   & ...        & G81 \\

$219$   &  $100$    &  ...    & ...   & ...  & ...   & 80\p16     & A85 \\
$94$    &  $70$     &  ...    & ...   & ...  & ...   & 80\p16     & A85 \\
$172$   &  $23$     &  34\p7  & ...   & ...  & ...   & 75\p15     & A85 \\
$125$   &  $-54$    &  37\p7  & ...   & ...  & ...   & 75\p15     & A85 \\
$77$    &  $-130$   &  36\p7  & ...   & ...  & ...   & 100\p20    & A85 \\
$  30$  &  $-207$   &  48\p10 & ...   & ...  & ...   & 80\p16     & A85 \\
$-18$   &  $-283$   &  31\p6  & ...   & ...  & ...   & 70\p14     & A85 \\
$59$    &  $-331$   &  ...    & ...   & ...  & ...   & 100\p20    & A85 \\
$-66$   &  $-360$   &  62\p8  & ...   & ...  & ...   & 70\p14     & A85 \\
$-113$  &  $-436$   &  ...    & ...   & ...  & ...   & 90\p18     & A85 \\
$-8$    &  $-203$   &  25\p7  & ...   & ...  & ...       & ...    & O89 \\

$-38$   & $-203$    &  24\p5  & ...   & ...  & ...       & ...    & O89 \\
$-18$   & $-233$    &  31\p7  & ...   & ...  & ...       & ...    & O89 \\
$-49$   & $-253$    &  30\p13 & ...   & ...  & ...       & ...    & O89 \\
$-77$   & $-253$    &  22\p25 & ...   & ...  & ...       & ...    & O89 \\
$-18$   & $-263$    &  24\p3  & ...   & ...  & ...       & ...    & O89 \\
$-28$   & $-263$    &  17\p2  & ...   & ...  & ...       & ...    & O89 \\
$-70$   & $-263$    &  27\p34 & ...   & ...  & ...       & ...    & O89 \\
$-58$   & $-323$    &  18\p3  & ...   & ...  & ...       & ...    & O89 \\
$-107$  & $-323$    &  3\p6  & ...   & ...  & ...        & ...   & O89 \\
$-91$   & $-347$    &  18\p3  & ...   & ...  & ...       & ...    & O89 \\
$-49$   & $-353$    &  13\p3  & ...   & ...  & ...       & ...    & O89 \\

$-106$  & $-335$    &  35\p2  & ...   & ...  & $64\pm4$    & ...       & H93 \\

$26$    & $-23$     &	$32_{-5}^{+30}$ &...  & ...  & ...  & ...   & C99 \\
$26$    & $-83$        &	$22_{-5}^{+30}$ &...  & ...  & ...    & ...       & C99 \\
$-13$   & $-163$       &	$22_{-5}^{+30}$ &...  & ...  & ...    & ...       & C99 \\

$144$   & $4$          &	$20_{-5}^{+30}$ &...  & ...  & ...    & ...       & C00 \\
$105$   & $-126$       &	$30_{-5}^{+30}$ &...  & ...  & ...    & ...       & C00 \\

\enddata
\tablenotetext{\dagger}{Rotational temperature measurements by
\citet{gus81} do not include an estimate of uncertainties.  For these
measurements, we assume an uncertainty of $\pm10$~K.}
\end{deluxetable}

\clearpage

\begin{figure}
\caption{{\it NOTE: Image too large for astro-ph.  Please download a full version of this paper at http://www.astro.columbia.edu/$\sim$herrnstein/NH3/paper/ .}{\it a)} Mean value of NH\3(1,1) main line opacity ($\ta$).
Contours are in integer steps.  {\it b)} Significance of opacity
determinations.  Contours are 1, 2, 3, 6, 9, and 12$\sigma$.  Hatches
denote pixels for which only a lower limit for $\ta$ could be
determined.  {\it c)} Mean value of intrinsic line width
($\dvi$). Typical intrinsic line widths are 15~\kms.  {\it d)}
Significance of intrinsic line width determinations.  Contours are in
steps of $3\sigma$.  Hatches denote pixels for which only an upper
limit for $\dvi$ could be determined.
\label{fig:dvtau} }
\end{figure}

\begin{figure}
\caption{{\it NOTE: Image too large for astro-ph.  Please download a full version of this paper at http://www.astro.columbia.edu/$\sim$herrnstein/NH3/paper/ .} {\it a)} Pseudo-color map of the derived (2,2)-to-(1,1)
rotational temperature.  Contours are in steps of 10~K.  {\it b)}
Significance of $T_{R21}$ determinations.  Contours are in steps of
3$\sigma$.  {\it c)} Derived values for the excitation temperature.
Contours are in steps of 4~K.  Compared to the rotational temperature,
the excitation temperature is smoothly distributed (especially in
regions with high S/N) with values between 4 and 8 K.  {\it d)}
Significance of the $T_{ex}$ values.  Contours are 3, 15, 30, and
45$\sigma$.
\label{fig:temps} }
\end{figure}

\begin{figure}
\plotone{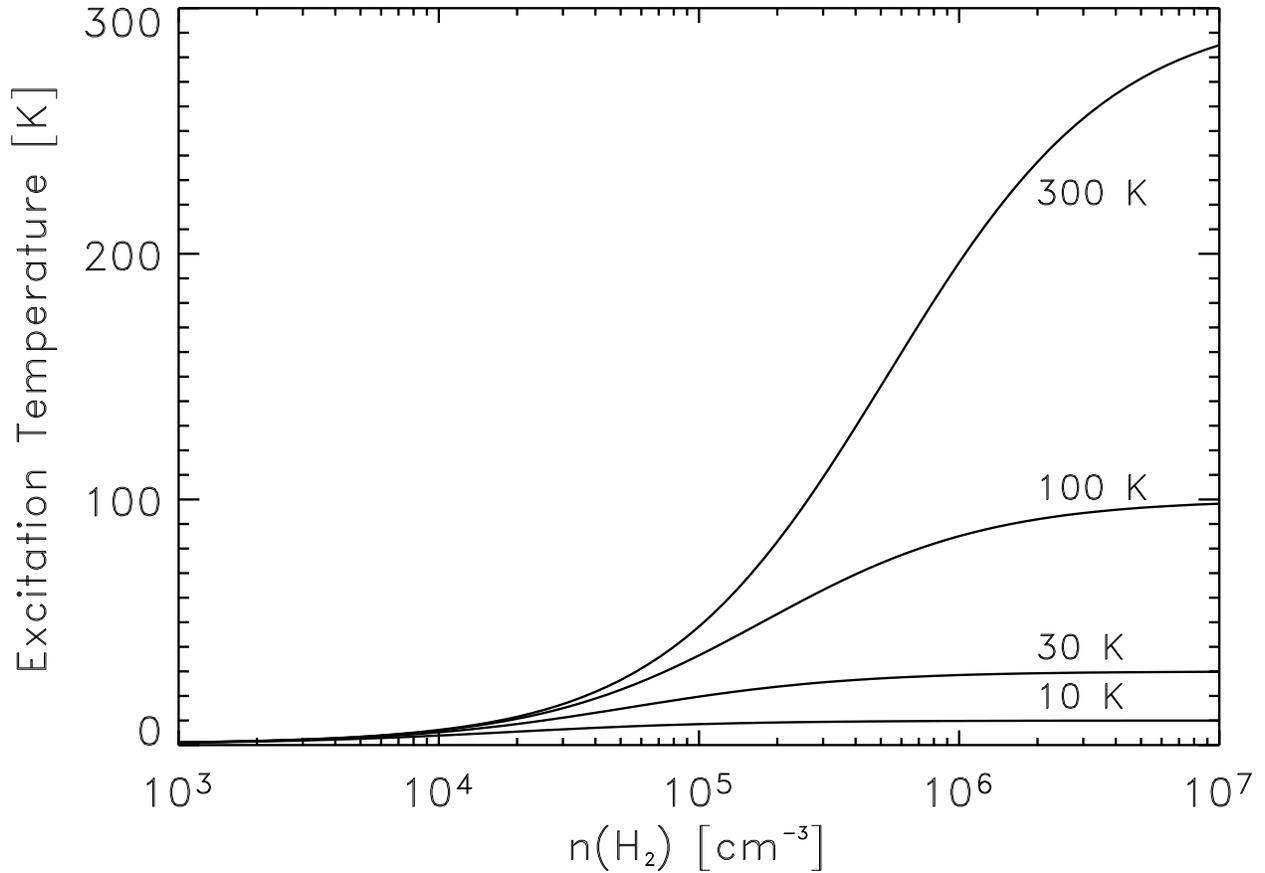}
\caption{ \label{fig:textk} Excitation temperature as a function of
H\2 density for kinetic temperatures of 10, 30, 100, and 300~K.  This
plot is valid for NH\3(1,1), (2,2), (3,3), and (6,6) because the
critical densities for these transitions are all equal to
$2\times10^3$~cm$^{-3}$.  Gas is not thermalized until $n(H_2)\approx
10^6$~cm$^{-3}$.}
\end{figure}

\begin{figure}
\caption{{\it NOTE: Image too large for astro-ph.  Please download a full version of this paper at http://www.astro.columbia.edu/$\sim$herrnstein/NH3/paper/ .} \label{fig:nnh3} Total column density of NH\3.  Contours are
in steps of $4\times10^{15}$~cm$^{-2}$.  Significance of $N_{{\rm NH_3}}$
determinations.  Contours are in steps of 1.5$\sigma$.  }
\end{figure}	

\begin{figure}
\plottwo{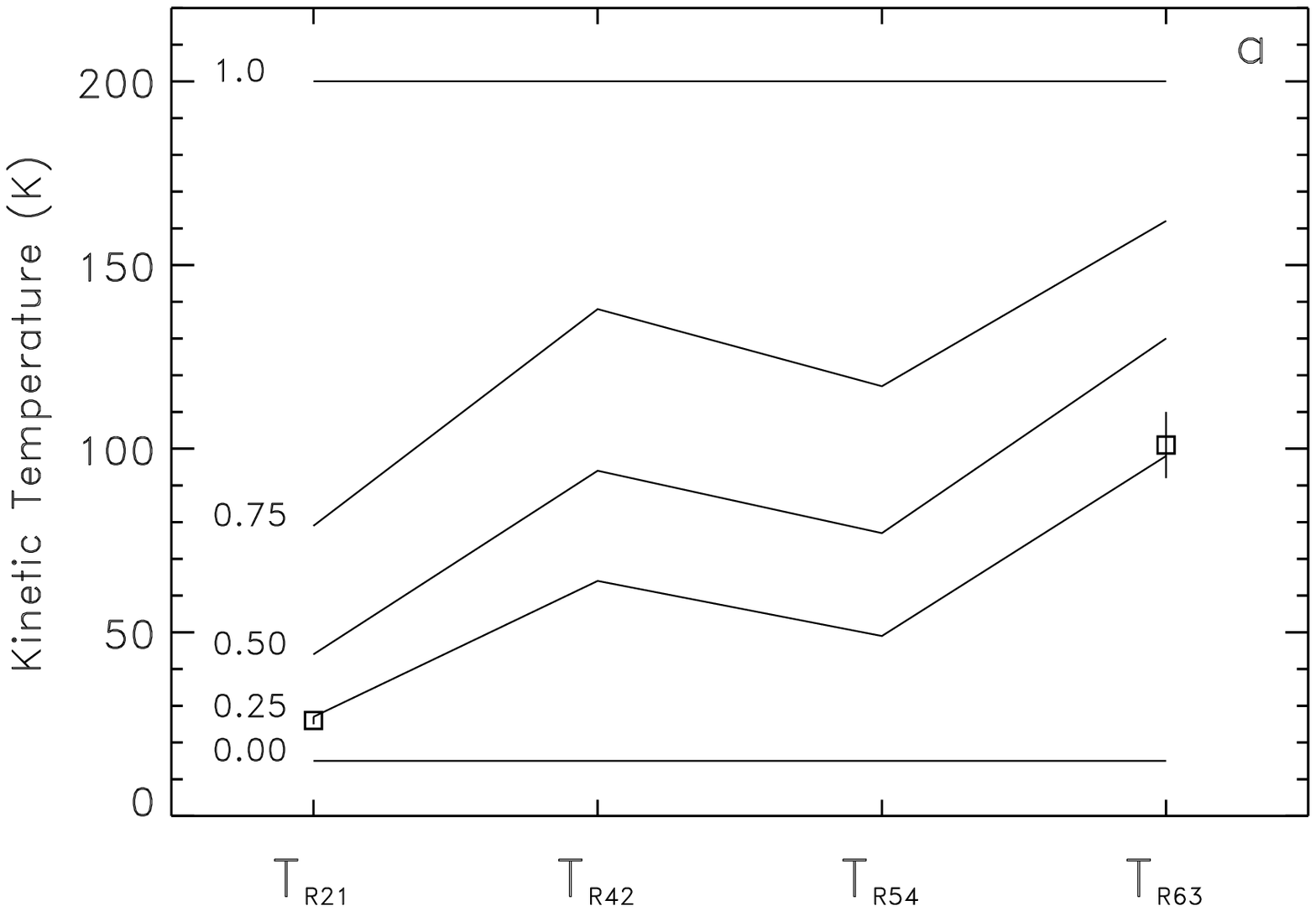}{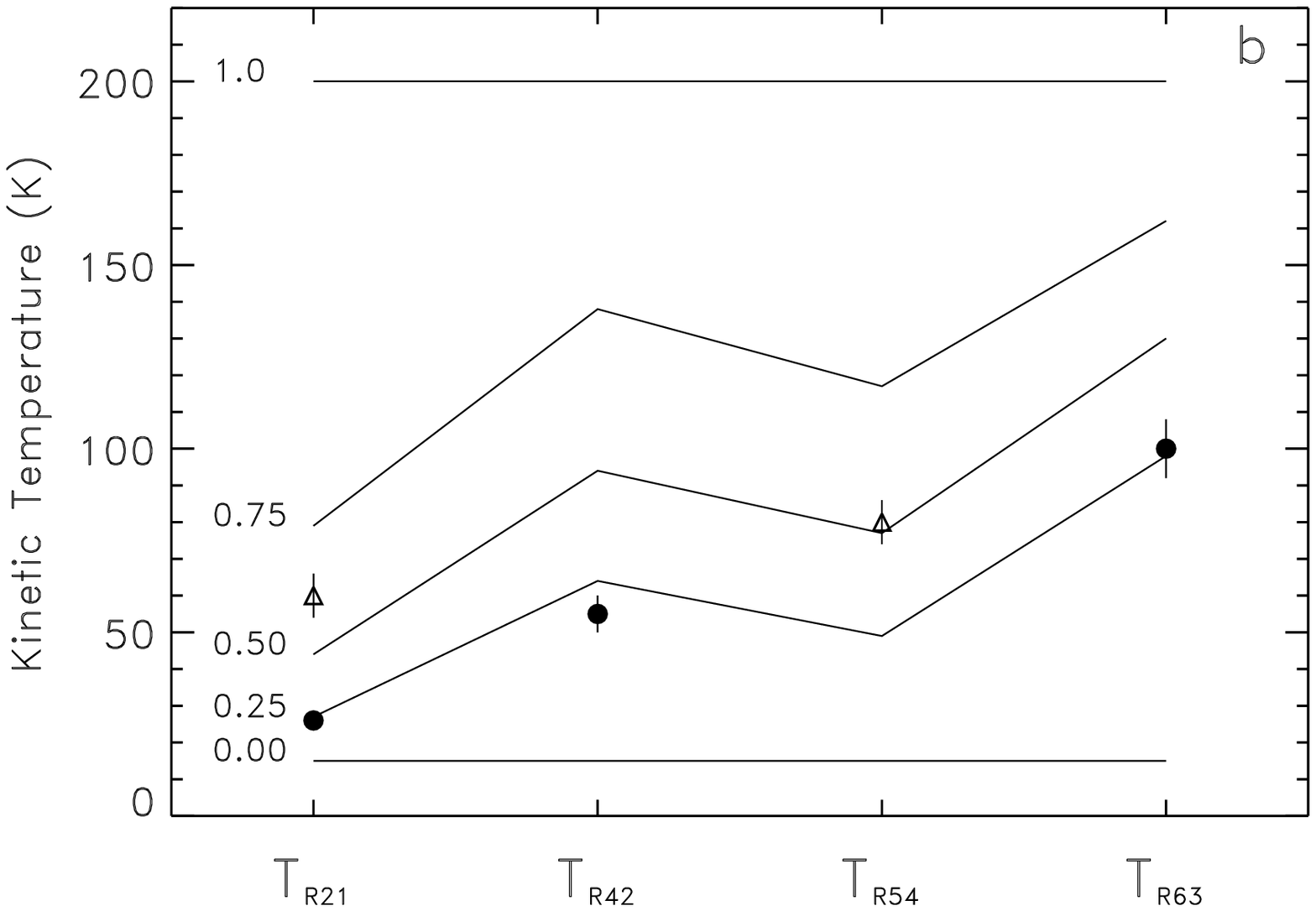}
\caption{ \label{fig:trtk} Calculated kinetic temperature for
$T_{R21}$, $T_{R42}$, $T_{R54}$, and $T_{R63}$ for six different
combinations of hot gas at 200~K and cold gas at 15~K.  {\it a)}
Weighted mean kinetic temperatures calculated from the $T_{R21}$ and
$T_{R63}$ measurements presented in this paper are overlaid as
squares.  {\it b)} Weighted mean kinetic temperatures calculated from
published $T_{R21}$, $T_{R42}$, and $T_{R63}$ measurements in Figure
\ref{fig:radial} are overlaid as filled circles.  The measurements of
$T_{R21}$ and $T_{R54}$ from \citet{hut93b} are overlaid as
triangles.}
\end{figure}

\begin{figure}
\caption{{\it NOTE: Image too large for astro-ph.  Please download a full version of this paper at http://www.astro.columbia.edu/$\sim$herrnstein/NH3/paper/ .} \label{fig:beams} {\it a)} Position of the single dish
measurements listed in Table \ref{table:t3} overlaid on a grey scale
image of 1.2~mm emission \citep{zyl98}.  Contours denote 10, 20, 40,
and 80\% of the peak flux.  Labels correspond to the list of
observations in Table \ref{table:t2}.  {\it b)} Position of
interferometric observations in the same region.  Primary beams of the
interferometer are plotted.  For \citet{oku89}, the synthesized beam
of the NMA is also overlaid at the positions where temperature
measurements were made. }
\end{figure}	

\begin{figure}
 \includegraphics[width=4.5in, angle=90]{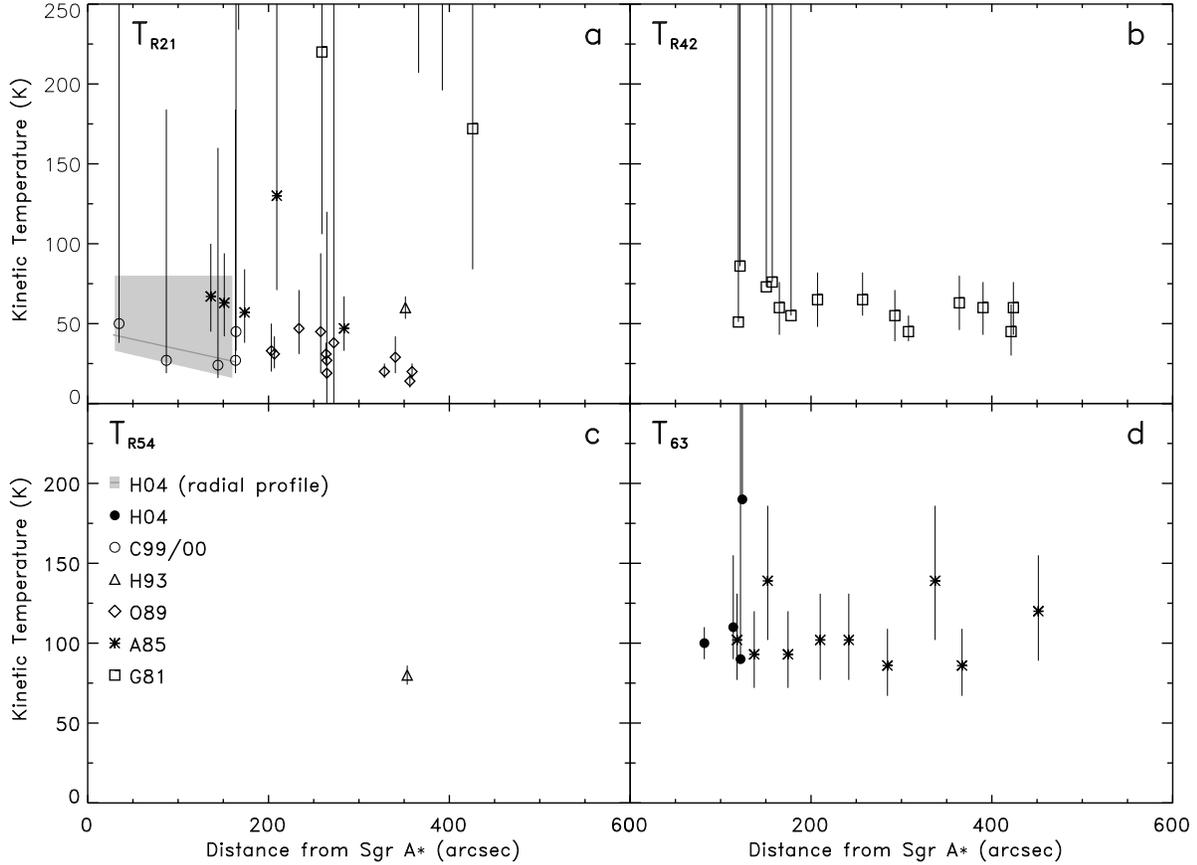}
\caption{\label{fig:radial} Measured kinetic temperature versus
distance from Sgr~A* calculated from the rotational temperatures
listed in Table \ref{table:t3} using \citet{dan88}.  The four panels
show kinetic temperatures derived from {\it a)} $T_{R21}$, {\it b)}
$T_{R42}$, {\it c)} $T_{R54}$, and {\it d)} $T_{R63}$.  Uncertainties
in the radial distance from Sgr~A*, which are equal to the FWHM of the
beam, have been omitted for clarity.  The dark grey line in panel $a$
shows the best fit radial profile for kinetic temperatures derived
from our data.  The surrounding light grey box shows the approximate
range of kinetic temperatures from pixels with a significance of
$\ge3\sigma$.}
\end{figure}

\begin{figure}
\includegraphics[width=5.0in, angle=270]{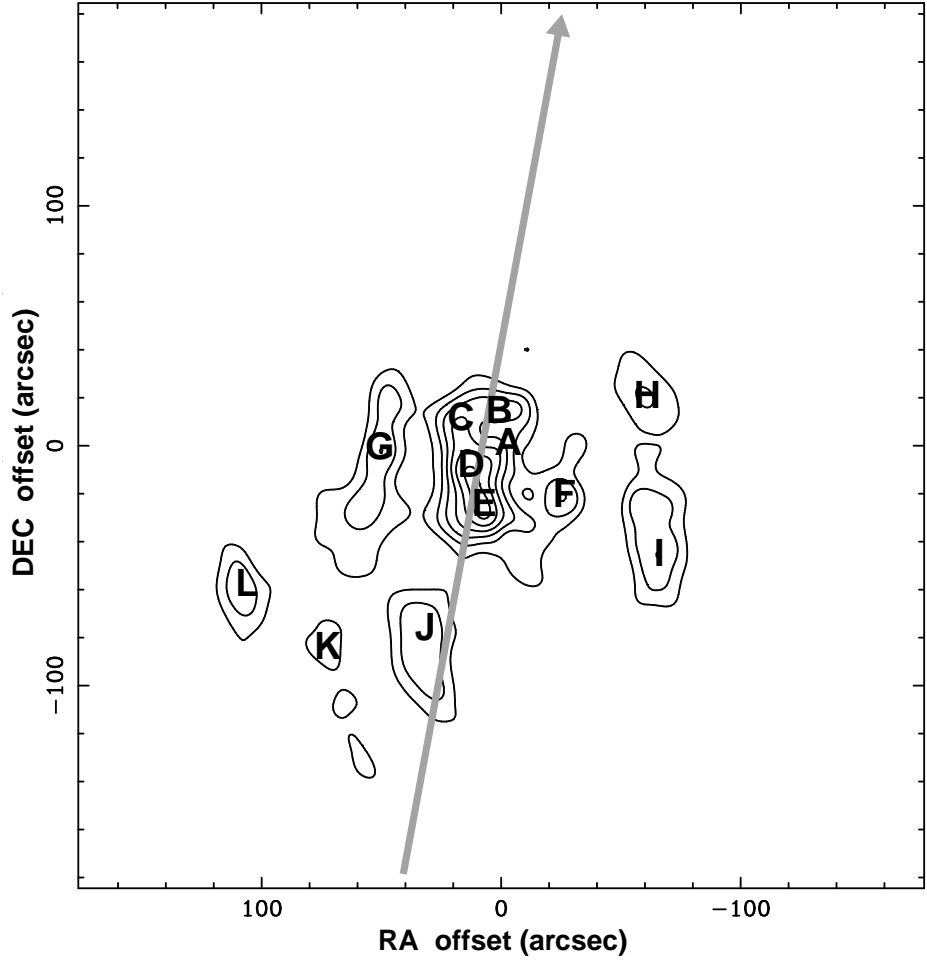} \singlespace 
\caption{
 \label{fig:pos} Position of twelve spectra and one position-velocity cut
 overlaid on a velocity integrated map of NH\3(6,6).  Sgr~A* is located
 at ($\Delta\alpha=0''$,$\Delta\delta=0''$) (spectrum A).} 
\end{figure}

\begin{figure}
\includegraphics[width=4.3in, angle=90]{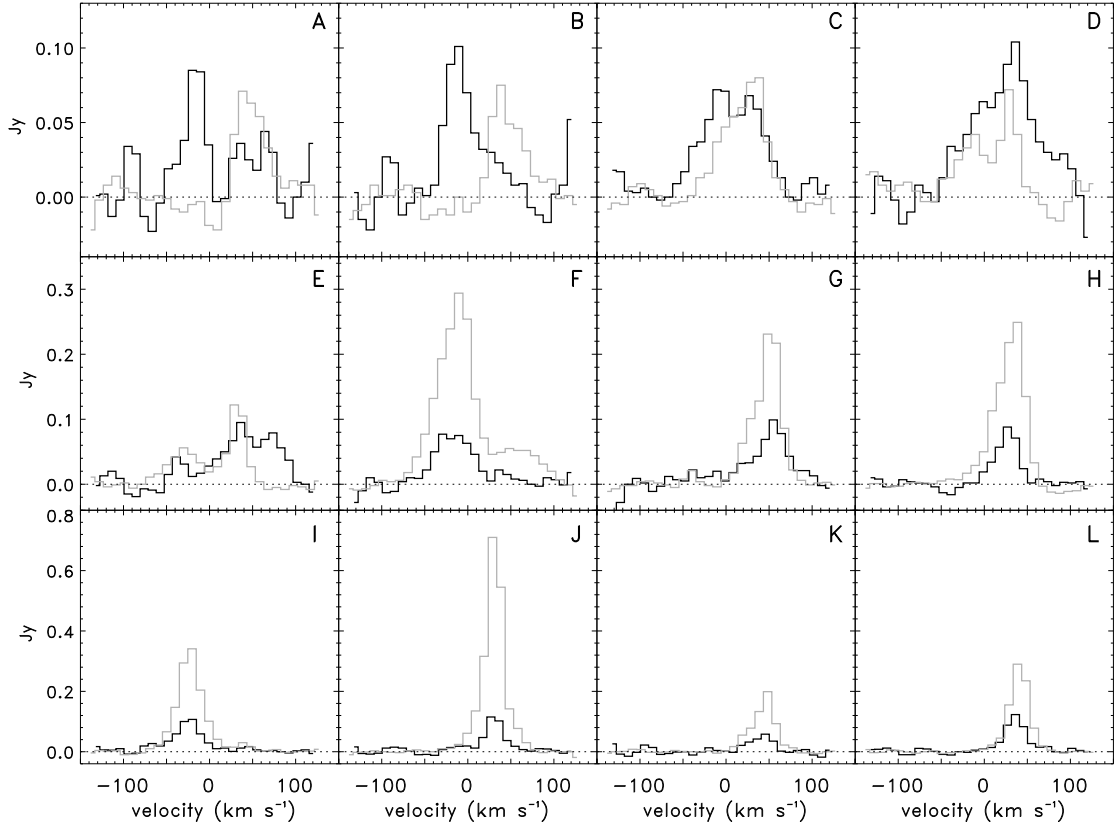}
\caption{ \label{fig:spec} Spectra from NH\3(3,3) (grey) and
NH\3(6,6) (black) at the positions labeled in Figure
\ref{fig:pos}.  For both transitions, we use image cubes deconvolved
using a $14''$ Gaussian taper.  Spectra are Hanning smoothed and the
noise levels are $\sigma^*_{33,ch}=7.6$~mJy~Beam\1 and
$\sigma^*_{66,ch}=9.5$~mJy~Beam\1.}
\end{figure}	

\begin{figure}
\plotone{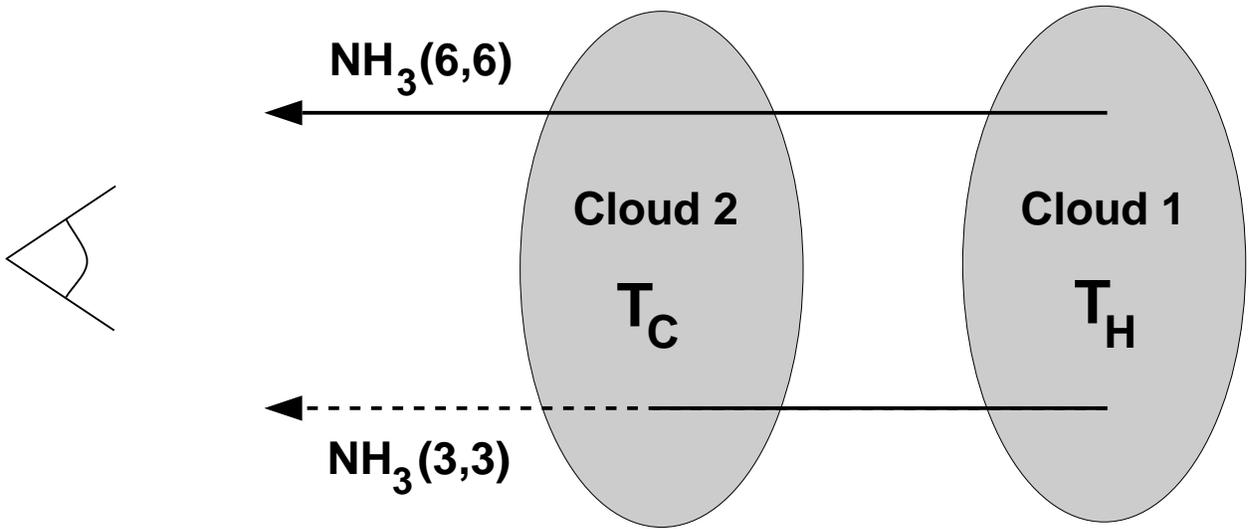}
\caption{ \label{fig:6633model} Schematic model of the orientation of
a hot and cold cloud that can produce ``un-physical'' NH\3
(6,6)-to-(3,3) line ratios.}
\end{figure}

\begin{figure}
\plottwo{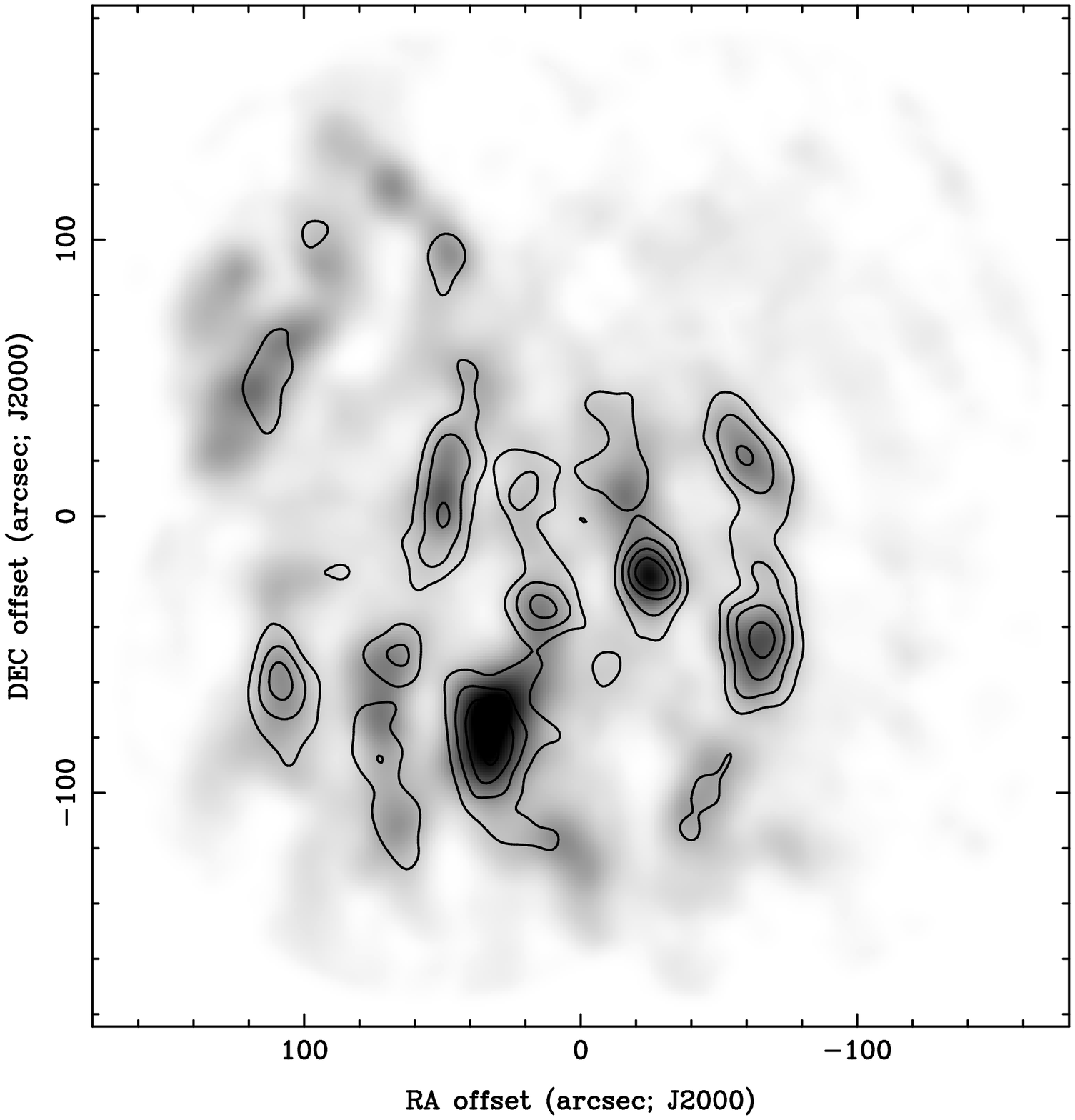}{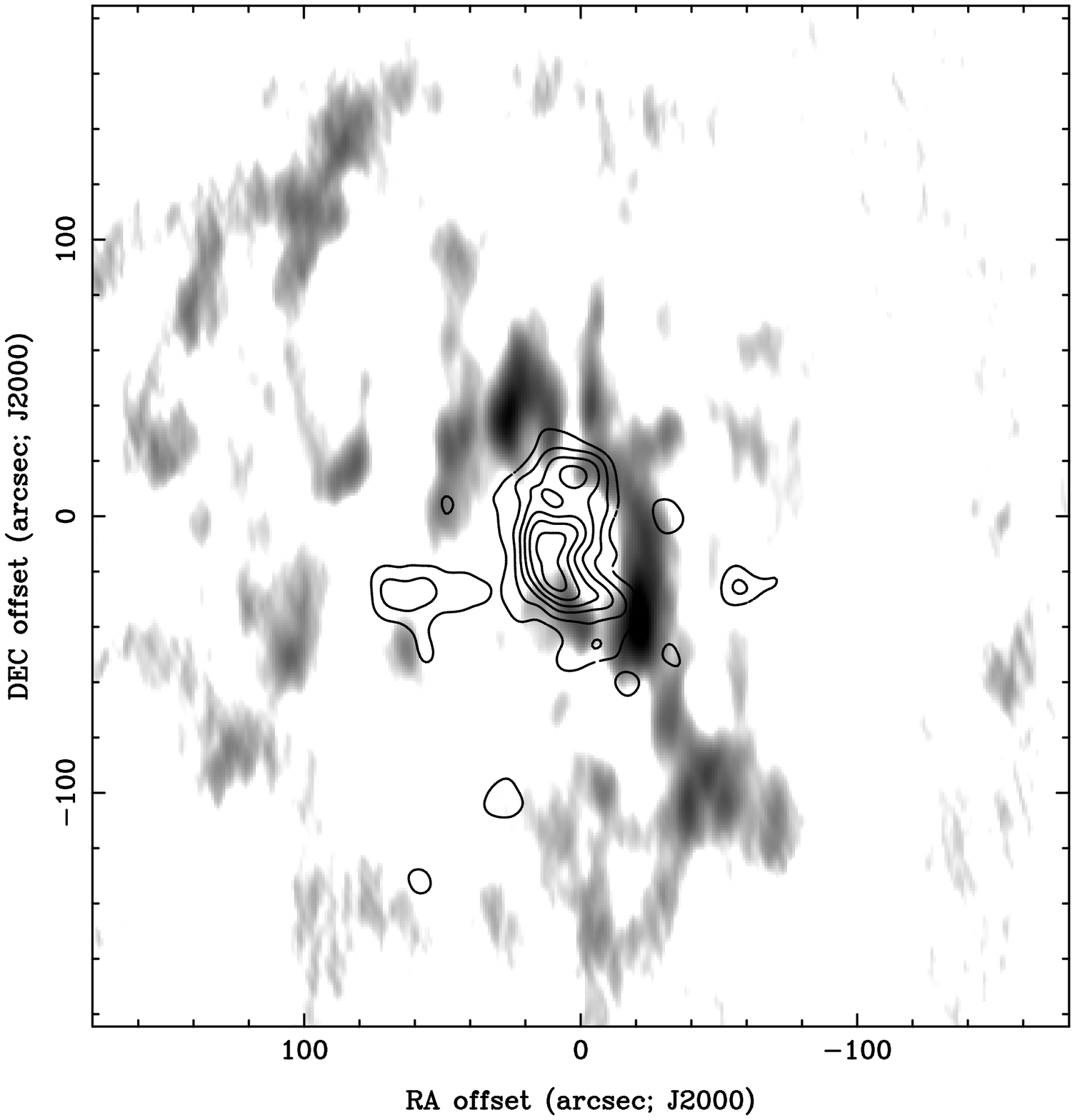}
\caption{\label{fig:f12} {\it Left:} Velocity integrated map of
NH\3(6,6) emission at all points with NH\3 (6,6)-to-(3,3) line ratios
less than one overlaid on a grey scale map of velocity integrated
NH\3(3,3) emission.  Contours are in steps of $3\sigma_{66}$ where
$\sigma_{66}=0.33$~Jy~Beam\1~km~s\1.  Low line ratio gas has a similar
distribution to velocity integrated emission from lower-energy
rotation inversion transitions.  {\it Right:} Velocity integrated map
of NH\3(6,6) emission at all points with NH\3 (6,6)-to-(3,3) line
ratios greater than one overlaid on a grey scale map of velocity
integrated HCN(1--0) emission \citep{wri01}.  Contour levels are the
same as in the left-hand panel.  High-line-ratio (HLR) gas is confined
to the inner 2~pc, interior to the CND, and a tongue of emission
extending to the east at $\Delta\delta=-30''$. }
\end{figure}

\begin{figure}
\plotone{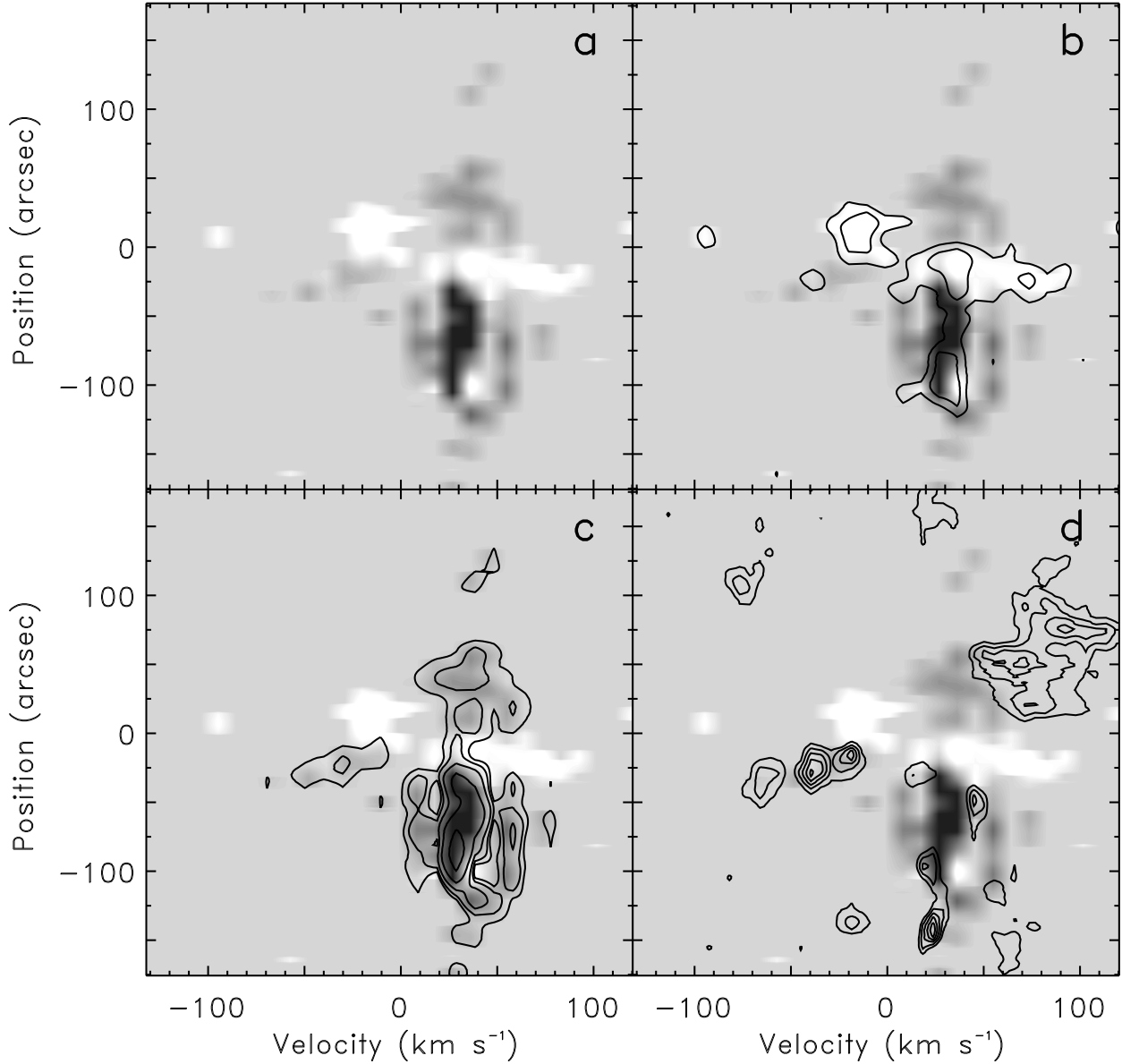} 
\caption{\label{fig:pv} Position velocity diagrams showing the
distribution of molecular gas in the central 2~pc.  The position of
the cut is labeled in Figure \ref{fig:pos}.  A position of $0''$
corresponds to $\Delta\delta=0''$, and positive positions are to the
northwest of Sgr~A*.  In all four panels, $S_\nu(6,6)-S_\nu(3,3)$ is
plotted in grey scale.  High-line-ratio gas ($S_\nu(6,6)>S_\nu(3,3)$)
appears as white, while low-line-ratio gas ($S_\nu(6,6)<S_\nu(3,3)$)
is dark grey.  Although $S_\nu(6,6)-S_\nu(3,3)$ ranges between --0.65
and 0.13, the grey scale stretch ranges from --0.25 to 0.25 to
highlight the different components in the data.  {\it b)}~NH\3(6,6)
overlaid in contour steps of $3\sigma_{66,ch}$.  NH\3(6,6) emission is
observed from both the HLR and LLR gas.  {\it c)}~NH\3(3,3) overlaid
in contour steps of 3, 6, 12, 24, and $48~\sigma_{33,ch}$.  NH\3(3,3)
emission comes only from LLR gas.  {\it d)}~HCN(1--0) from
\citet{wri01} overlaid in contour steps of 0.25~Jy~Beam\1.  The lack
of HCN(1--0) emission at 30~km~s\1 is the result of self-absorption.}
\end{figure}

\clearpage
\begin{figure}
\epsscale{0.6}
\plotone{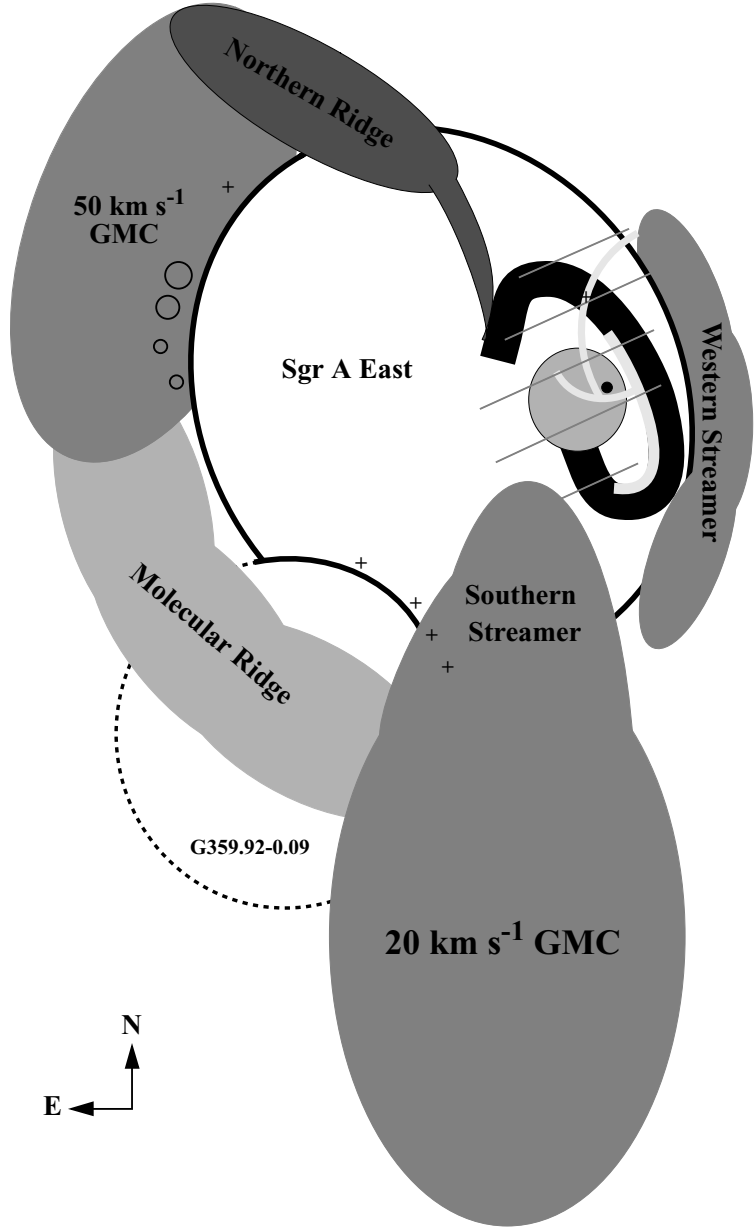}
\caption{ \label{fig:f17} Schematic drawing of the Galactic center as
seen in the plane of the sky.  }
\end{figure}

\begin{figure}
\epsscale{0.6}
\rotate
\includegraphics[width=4.75in,angle=270]{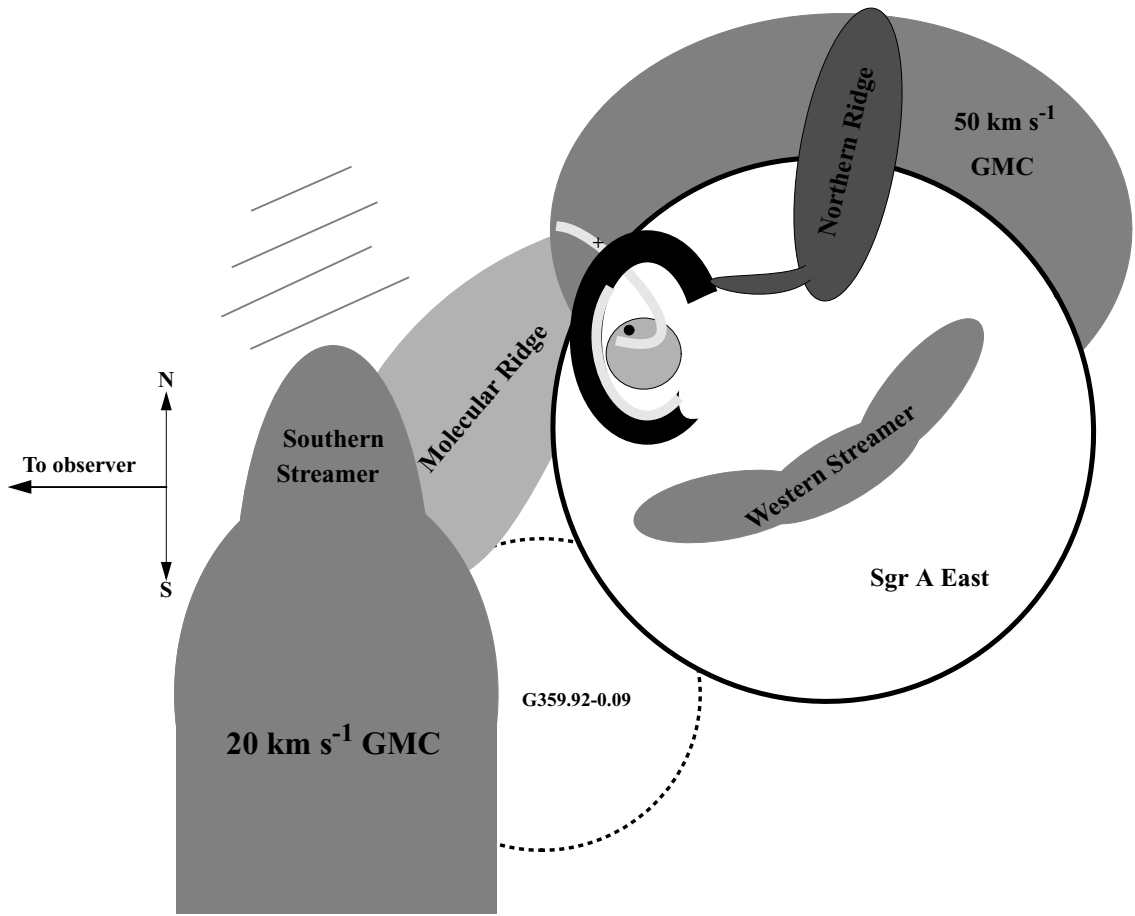}
\caption{ \label{fig:f18} Schematic drawing of the Galactic center
with left denoting the direction to the observer and the vertical
direction denoting north and south.  }
\end{figure}

\begin{figure}
\epsscale{0.6} \rotate
\includegraphics[width=4.75in,angle=270]{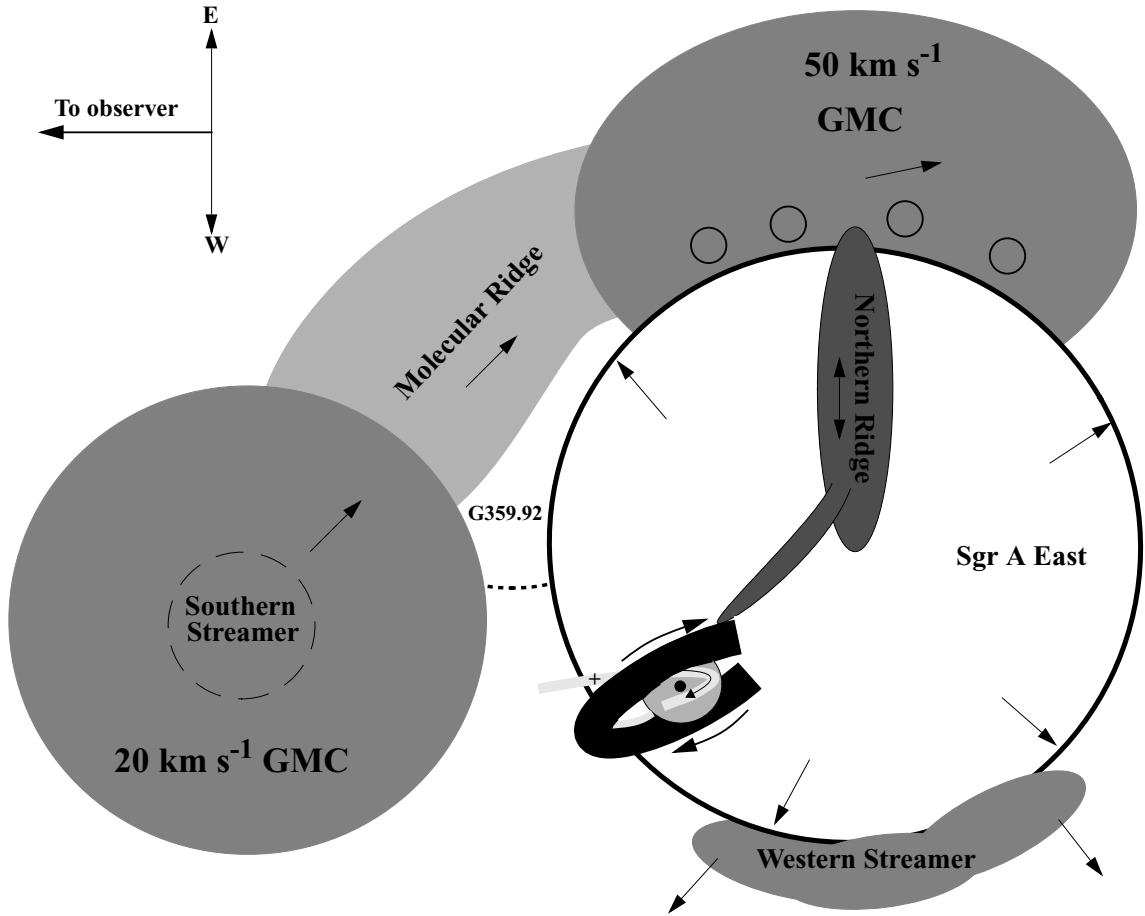}
\caption{ \label{fig:f19} Schematic drawing of the Galactic center
with left denoting the direction to the observer and the vertical
direction denoting east and west.  Arrows show the inferred motions of
the main features based on the Doppler-shifted velocities.}
\end{figure}

\end{document}